\documentclass[acmsmall,nonacm]{acmart}
\usepackage{acronym}
\usepackage{subfigure}
\usepackage{xspace}
\usepackage{booktabs}
\usepackage{tikz}
\usepackage{moreverb}
\usepackage{amsmath}
\usepackage{listings}
\usepackage{algorithm}
\usepackage{enumitem}
\usepackage[noend]{algpseudocode}

\newcommand{\tr}[1]{\ensuremath{#1}^\intercal} 

\acrodef{AB}{Asymptotic Bounds}
\acrodef{BAS}{Blocking After Service}
\acrodef{CB}{Composite Bounds}
\acrodef{BSB}{Balanced System Bounds}
\acrodef{CMVA}{Conditional MVA}
\acrodef{CTMC}{Continuous Time Markov Chain}
\acrodef{DTMC}{Discrete Time Markov Chain}
\acrodef{FCFS}{First-Come First-Served}
\acrodef{GB}{Geometric Bounds}
\acrodef{IS}{Infinite Server}
\acrodef{JMT}{Java Modeling Tools}
\acrodef{LCFS-PR}{Last-Came First-Served, Preemptive Resume}
\acrodef{LCFS}{Last-Come-First-Served}
\acrodef{MVA}{Mean Value Analysis}
\acrodef{MTBF}{Mean Time Between Failures}
\acrodef{MTTA}{Mean Time To Absorption}
\acrodef{MTTF}{Mean Time To Failure}
\acrodef{PDF}{Probability Density Function}
\acrodef{PFQN}{Product-form Queueing Network}
\acrodef{PMF}{Probability Mass Function}
\acrodef{PS}{Processor Sharing}
\acrodef{QN}{Queueing Network}
\acrodef{RS-RD}{Repetitive Service with Random Destination}
\acrodef{SIRO}{Service In Random Order}

\newcommand{\queueing}{\texttt{queueing}\xspace}

\begin{document}

\title{A Software Package for Queueing Networks and Markov Chains analysis}

\author{Moreno Marzolla}
\email{moreno.marzolla@unibo.it}
\orcid{0000-0002-2151-5287}
\affiliation{
  \institution{Universit\`a di Bologna}
  \department{Department of Computer Science and Engineering (DISI)}
  \streetaddress{Mura Anteo Zamboni 7}
  \city{Bologna}
  \postcode{I-40126}
  \country{Italy}
}
\email{moreno.marzolla@unibo.it}

\begin{abstract}
  Queueing networks and Markov chains are widely used for conducting
  performance and reliability studies. In this paper we describe the
  \queueing package, a free software package for queueing networks and
  Markov chain analysis for GNU Octave. The \queueing package provides
  implementations of numerical algorithms for computing transient and
  steady-state performance measures of discrete and continuous Markov
  chains, and for steady-state analysis of single-station queueing
  systems and queueing networks. We illustrate the design principles
  of the \queueing package, describe its most salient features and
  provide some usage examples.
\end{abstract}



\maketitle

\section{Introduction}

\acp{QN} and Markov chains are powerful modeling notations that are
commonly used for capacity planning, bottleneck analysis and
performance evaluation of systems~\cite{bolch}. Analyzing~\acp{QN} and
Markov chains involves the computation of metric such as the system
throughput of a~\ac{QN}, or the stationary state occupancy
probabilities of a Markov chain. Symbolic, numerical, and
simulation-based techniques have been developed to compute these
metrics.

In this paper we describe the \queueing package for GNU Octave, a free
environment for numerical computing~\cite{octave}. The \queueing
package provides implementations of numerical algorithms for
(\emph{i})~transient and stationary analysis of discrete and
continuous Markov chains; (\emph{ii})~stationary analysis of
single-station queueing systems; (\emph{iii})~stationary analysis of
some classes of product-form~\acl{QN}.

\begin{table}
  \small
  \begin{tabular}{lp{.65\textwidth}l}
    \toprule
    {\bf Name} & {\bf Description} & {\bf License} \\
    \midrule
    \href{http://jmt.sourceforge.net/}{JMT}~\cite{jmt} & Java tool for workload characterization, simulation-based queueing network modeling & GNU GPL\\
    \href{http://line-solver.sourceforge.net/}{LINE}~\cite{Casale:2019} & Performance and reliability analysis based on queueing models & BSD-3 \\
    \href{http://www.perfdynamics.com/Tools/PDQcode.html}{PDQ}~\cite{PDQ} & queueing networks & MIT\\
    \href{https://cran.r-project.org/package=queueing}{queueing}~\cite{queueing-r} & R package for analyzing single-station queueing systems and queueing networks & GNU GPL\\
    \href{https://sharpe.pratt.duke.edu/}{SHARPE}~\cite{sharpe} & reliability modeling, Markov and semi-Markov models, Petri nets, queueing networks & Proprietary\\
    \bottomrule
  \end{tabular}
  \caption{Some software package for queueing network analysis that
    are currently available and actively
    maintained.}\label{tab:packages}
\end{table}

Although~\acp{QN} and~Markov chains are well studied topics,
relatively few computer implementations of solution algorithms are
available and actively
supported~\cite{perfhist}. Table~\ref{tab:packages} lists some
software tools that are relevant for this paper.

JMT~\cite{jmt} is a Java package for workload characterization,
bottleneck analysis, and~\ac{QN} modeling. JMT has a GUI that
simplifies the definition and analysis of~\ac{QN} models, although it
can also be used from the command line. JMT uses a simulation engine
as its main solution technique, so it can support extended features
(non-Markovian queues, fork/join systems, passive resources, and
others) that are difficult if not impossible to handle numerically.

The LINE solver~\cite{Casale:2019} is a free MATLAB toolbox for
analyzing extended and layered queueing
networks~\cite{lqn}. Extended~\acp{QN} support features, such as
simultaneous resource possession, fork/join systems, finite capacity
regions and others, allowing more accurate models to be
defined. This comes with the drawback that extended~\acp{QN} are more
difficult to analyze numerically. LINE can delegate the solution of
these models to external solvers such as~JMT.

PDQ~\cite{PDQ} is an implementation of the~\ac{MVA} algorithm for
closed, single-class networks. PDQ provides bindings for different
programming languages: at the time of writing, C, Perl, Python and R
are supported.

The queueing package for R~\cite{queueing-r} (that, despite the name,
is unrelated to the software described in this paper) is a free
package for analyzing product-form~\acp{QN} written in the R
language~\cite{r-lang}. It supports product-form open and closed,
single and multiclass networks.

SHARPE (Symbolic Hierarchical Automated Reliability and Performance
Evaluator)~\cite{sharpe} is a hierarchical modeling tool that supports
any combination of different types of performance and reliability
models (product-form queueing networks, Petri nets, Markov chains,
fault trees). SHARPE has both a command-line and a graphical
interface, and has been under development since the early 80s. It is
the only tool of those reviewed that has a non-free license.

The \queueing package presented in this paper is somewhat orthogonal
to above tools, in the sense that it has been developed around
specific design goals which are only partially considered by other
packages; of course, this implies that it has some limitations which
might be addressed by other tools.

One of the design goals of the \queueing package is to provide
reference implementations of some fundamental ``textbook'' algorithms
for~\ac{QN} and Markov chain analysis, like other research communities
are doing since a long time (e.g., linear algebra algorithms). To this
aim, efficiency has sometimes been sacrificed in favor of code
readability. The availability of reference implementations is useful
also for teaching purposes: students can immediately put the textbook
algorithms at work to solve practical problems, encouraging ``learning
by doing''. The author is aware of several Universities that are using
the \queueing package to teach performance modeling classes.

The GNU Octave language, being a large subset of the MATLAB language,
is well suited for implementing numerical algorithms that operate on
arrays and matrices in a concise and understandable way. Moreover, it
allows complex performance studies can be done quickly, since models
involving repetitive or embedded structure can be defined
programmatically. Parametric model evaluation or ad-hoc analyses are
also possible. The \queueing package has been contributed to the
Octave-forge public repository (\url{https://octave.sourceforge.io/}).
This means that \queueing can be easily installed from the Octave
prompt using the standard command \verb+pkg install+.

Any design decision inevitably carries some drawbacks. The GNU Octave
environment allows a great degree of flexibility, but imposes a steep
learning curve that might deter the occasional user. The focus on
well-known classic algorithms neglects more recent results or less
frequently used techniques. Yet, both issues can be addressed.  A more
comfortable interface, e.g., a GUI, can be built either as an
independent application, or by leveraging existing tools (e.g., JMT)
and then delegating the computations to the \queueing package. More
algorithms can be implemented and contributed for inclusion in the
\queueing package, that is free software and as such can be extended
by anyone. A few contributors already did so.

This paper is structured as follows. In Section~\ref{sec:design} we
illustrate the design principles behind the~\queueing package. The
next sections are devoted to illustrate the functions for analyzing
Markov chains (Section~\ref{sec:markov-chains}), single-station
queueing systems (Section~\ref{sec:single-station}) and queueing
networks (Section~\ref{sec:queueing-networks}). The presentation
focuses on the features provided, rather than the algorithmic details;
comprehensive bibliographic references are provided for the interested
reader. Although this paper is not intended to be a substitute of the
package user's manual, a few examples will be shown to better
illustrate its use. Finally, concluding remarks are given in
Section~\ref{sec:conclusions}.

\section{Design Principles}\label{sec:design}

The~\queueing package is a collection of functions for computing
transient and steady-state performance measures of queueing networks
and Markov chains. It has been under development over the last decade
to support the author's research and teaching activity in the area of
performance modeling of systems. The~\queueing package consists of a
set of m-files written in the GNU Octave~\cite{octave} dialect of the
MATLAB programming language; therefore, \queueing does not require any
special installation procedure, nor does it require a compiler to
generate executable code.

The decision of targeting GNU Octave was made at the beginning of the
development effort. GNU Octave started its existence as a free MATLAB
clone, but ended up providing extensions and additional features, some
of which have been exploited by \queueing (see below). More
importantly, GNU Octave is free software and runs on all major
operating systems, so it does not represent an entry barrier for
potential users.

GNU Octave supports most of the standard MATLAB syntax, plus some
extensions. For example, \verb+!+ can be used as the logical
\emph{not} operator; structured blocks such as the \verb+if+ and
\verb+for+ constructs can be terminated with the \verb+endif+ and
\verb+endfor+ keywords, respectively, to improve readability. The
Texinfo markup notation~\cite{texinfo} can be used for the
documentation text embedded in function files. This feature has been
extensively used: the documentation of each function in the \queueing
package can be displayed using the \verb+help+ command during
interactive sessions. The user's manual, in~PDF and~HTML formats, is
built from the Texinfo documentation extracted from the source
files. This guarantees that the user's manual is always consistent
with the help text.

\paragraph{Naming conventions}
Most of the functions in the \queueing package obey a common naming
convention. Function names are the concatenation of several parts,
beginning with a prefix that indicates the class of problems the
function addresses:\medskip

\begin{tabular}{p{15mm}l}
\verb+ctmc-+ & Functions dealing with continuous-time Markov chains\\
\verb+dtmc-+ & Functions dealing with discrete-time Markov chains\\
\verb+qs-+   & Functions dealing with single-station queueing systems\\
\verb+qn-+   & Functions dealing with queueing networks
\end{tabular}\medskip

Functions that handle Markov chains (Section~\ref{sec:markov-chains})
start with either the \verb+ctmc+ or \verb+dtmc+ prefix, that may be
followed by a string that hints at what the function does:\medskip

\begin{tabular}{p{15mm}l}
\verb+-bd+    & Birth-Death process\\
\verb+-mtta+ & Mean Time to Absorption\\
\verb+-fpt+  & First Passage Times\\
\verb+-exps+ & Expected Sojourn Times\\
\verb+-taexps+ & Time-Averaged Expected Sojourn Times
\end{tabular}\medskip

Therefore, function \verb+ctmcbd+ returns the infinitesimal generator
matrix for a continuous birth-death process, while \verb+dtmcbd+
returns the transition probability matrix for a discrete birth-death
process. Functions \verb+ctmc+ and \verb+dtmc+ (without any suffix)
compute steady-state and transient state occupancy probabilities
for~\acp{CTMC} and~\acp{DTMC}, respectively.

Functions whose name starts with \verb+qs-+ deal with single station
queueing systems (Section~\ref{sec:single-station}). The suffix
describes the type of system, e.g., \verb+qsmm1+ for $M/M/1$,
\verb+qnmmm+ for $M/M/m$ and so on.

Finally, functions whose name starts with \verb+qn-+ deal with
queueing networks (Section~\ref{sec:queueing-networks}). The character
that follows indicates the type of network (\verb+o+~=~open network,
\verb+c+~=~closed network), and whether there is a single (\verb+s+)
or multiple (\verb+m+) customer classes.\medskip

\begin{tabular}{p{15mm}l}
\verb+-os-+ & Open, single-class network\\
\verb+-om-+ & Open, multiclass network\\
\verb+-cs-+ & Closed, single-class network\\
\verb+-cm-+ & Closed, multiclass network\\
\verb+-mix-+ & Mixed network with open and closed classes of customers
\end{tabular}\medskip

The last part of the function name indicates what the function computes:\medskip

\begin{tabular}{p{15mm}l}
\verb+-aba+ & Asymptotic Bounds\\
\verb+-bsb+ & Balanced System Bounds\\
\verb+-gb+ & Geometric Bounds\\
\verb+-pb+ & PB Bounds\\
\verb+-cb+ & Composite Bounds\\
\verb+-mva+ & Mean Value Analysis (MVA)\\
\verb+-cmva+ & Conditional MVA\\
\verb+-mvald+ & MVA with load-dependent servers\\
\verb+-mvabs+ & Approximate~MVA using Bard and Schweitzer's approximation\\
\verb+-mvablo+ & Approximate~MVA for blocking queueing networks\\
\verb+-conv+ & Convolution algorithm\\
\verb+-convld+ & Convolution algorithm with load-dependent servers
\end{tabular}\medskip

\paragraph{Validation}
One important issue of numerical software is to make sure that the
computed results are correct. Almost all functions in the \queueing
package include unit tests embedded as specially-formatted comments
inside the source code. The unit tests are used to check the results
against reference values from the literature. When reference results
are not available, cross-validation with the output of different
functions on the same model (if available), or with the output of
other packages have been used. For example, a closed product-form
network can be analyzed by~\ac{MVA} or using the convolution
algorithm; therefore it is possible to apply the functions
\verb+qncsmva()+ and \verb+qncsconv()+ on the same model and check
whether their results agree up to known numerical
problems~\cite{cmva}. Results have also been compared with those
produced by different tools. This was helpful to investigate an issue
with the \verb+qncmmva()+ function, whose result on the model
described in~\cite[Figure 7, p. 9]{schwetman-testing} did not agree
with the one reported in that paper. The model was analyzed
with~\ac{JMT} that confirmed the values computed by the \queueing
package.

\section{Markov chains}\label{sec:markov-chains}

A stochastic process is a set of random variables $\{X(t),\ t \in T\}$
where each~$X(t)$ is indexed by a time parameter $t \in T$.  The state
space is the set of all possible values of~$X(t)$. A time-homogeneous
Markov chain is a stochastic process over the discrete state space
$\{1, \ldots, N\}$ for some given~$N$. In a~\ac{DTMC}, the time
parameter~$t$ assumes the discrete values in $T = \{0, 1, \ldots\}$,
while in a~\ac{CTMC} the time parameter assumes values in $T =
\left[0, +\infty\right)$.

In a time-homogeneous~\ac{DTMC} the conditional probability $p_{i,j} =
\Pr\{X(n+1) = j\ |\ X(n) = i\}$ that the system is in state~$j$ at
time $n+1$, given that the system was in state~$i$ at time~$n$, is
independent from~$n$, so that we can define a~\ac{DTMC} as a
stochastic matrix $\mathbf{P} \in \mathbb{R}^{N \times N}$, where
$p_{i,j} = \Pr\{X(n+1) = j\ |\ X(n) = i\}$ is the transition
probability from state~$i$ to state~$j$, $i \neq j$.

Similarly, in a time-homogeneous~\ac{CTMC} the conditional probability
$p_{i,j}(u, v) = \Pr\{X(v) = j\ |\ X(u) = i\}$, $v \geq u$, only
depends on the time difference $t = v - u$, and not on the specific
values of~$u$ and~$v$, so that we have $p_{i,j}(t) = \Pr\{X(u+t) =
j\ |\ X(u) = i\} = \Pr\{X(t) = j\ |\ X(0) = i\}$ for each $t \geq
0$. The evolution of a~\ac{CTMC} is defined by a generator matrix
$\mathbf{Q} \in \mathbb{R}^{N \times N}$ where~$q_{i,j}$ is the
transition rate from state~$i$ to state $j \neq i$. The diagonal
elements $q_{ii}$ are defined in such a way that the sum of each row
is zero, i.e., $\mathbf{1} \mathbf{Q} = \mathbf{0}$ ($\mathbf{1}$ and
$\mathbf{0}$ denote suitably sized row vectors of~$1$ and~$0$,
respectively).

Let~$\pi_i(0)$ be the probability that the system is in state~$i$ at
time~$0$. It can be shown that the state occupancy probabilities
$\boldsymbol{\pi}(t) = \left( \pi_1(t), \ldots, \pi_N(t)\right)$ at
time~$t$ can be computed as:

\begin{align}
  \boldsymbol{\pi}(n) &= \boldsymbol{\pi}(0) \mathbf{P}^n \quad \text{(DTMC)} &
  \boldsymbol{\pi}(t) &= \boldsymbol{\pi}(0) e^{\mathbf{Q}t} \quad \text{(CTMC)}
  \label{eq:mc:transient}
\end{align}

Under certain conditions~\cite{bolch} a Markov chain has a unique
stationary distribution $\boldsymbol{\pi}$ that is independent from
the initial state. The stationary distribution can be computed by
solving the linear systems:

\begin{align}
  \begin{cases}
    \boldsymbol{\pi} \mathbf{P} &= \boldsymbol{\pi} \\
    \boldsymbol{\pi} \tr{\mathbf{1}} &= 1
  \end{cases} \quad\text{(DTMC)} &&
  \begin{cases}
    \boldsymbol{\pi} \mathbf{Q} &= \boldsymbol{0} \\
    \boldsymbol{\pi} \tr{\mathbf{1}} &= 1
  \end{cases} \quad\text{(CTMC)}
  \label{eq:mc:stationary}
\end{align}

\noindent where $\tr{\mathbf{1}}$ is a column vector of~$1$.

Functions \verb+dtmc()+ and \verb+ctmc()+ compute the transient or
stationary state occupancy probabilities of a~\ac{DTMC} and~\ac{CTMC},
respectively, using a direct implementations of
equations~\eqref{eq:mc:transient} and~\eqref{eq:mc:stationary},
respectively. For example, the expression \verb+pn = dtmc(P,n,p0)+
computes the state occupancy probability vector \verb+pn+ after~$n$
steps of a~\ac{DTMC} with stochastic matrix~$P$ and initial state
probabilities \verb+p0+. If invoked with a single parameter as in
\verb+p = dtmc(P)+, the function computes the stationary state
distribution vector \verb+p+. \verb+ctmc()+ can be used in a similar
way to analyze~\acp{CTMC}.

The \queueing package provides other functions that compute metrics
used in reliability and performability studies. The~\ac{MTTA} of
a~\ac{DTMC} is defined as the average number of transitions required
to reach an absorbing state, given the initial occupancy probability
vector $\boldsymbol{\pi}(0)$ (a state is \emph{absorbing} if it has no
outgoing transitions). The~\ac{MTTA} can be computed from the
fundamental matrix $\mathbf{N} = (\mathbf{I} - \mathbf{P}_t)^{-1}$,
where~$\mathbf{P}_t$ is the restriction of the transition
matrix~$\mathbf{P}$ to transient states only, and~$\mathbf{I}$ is a
suitably sized square identity matrix. Given initial state occupancy
probabilities $\boldsymbol{\pi}(0)$, the mean number of steps
before entering any absorbing state is:
\begin{equation*}
\textit{MTTA} = \boldsymbol{\pi}(0)\tr{(\mathbf{1} \mathbf{N})}
\end{equation*}

Other metrics of interest include the \emph{first passage time}
$M_{i,j}$, defined as the average number of transitions before
state~$j$ is visited for the first time, starting from
state~$i$. Finally, the \emph{mean sojourn time} $L_i(n)$ is the
expected number of visits to state~$i$ during the first~$n$
transitions, for given initial state occupancy probabilities; the
ratio $L_i(n)/n$ is called \emph{time-averaged mean sojourn time}. All
these concepts can be easily defined for continuous-time Markov chains
as well.

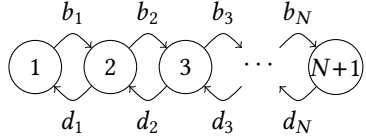
\begin{figure}
\begin{tikzpicture}
\tikzstyle{every circle node} = [minimum size=7mm,draw,inner sep=0pt];
\path
(0,0) node[circle] (a) {$1$}
(1,0) node[circle] (b) {$2$}
(2,0) node[circle] (c) {$3$}
(3,0) node[circle,draw=white] (d) {$\cdots$}
(4,0) node[circle] (e) {$N\! +\! 1$};
\draw[->] (a) .. controls +(.5,.5) .. (b) node[midway,above] {$b_1$};
\draw[->] (b) .. controls +(-.5,-.5) .. (a) node[midway,below] {$d_1$};
\draw[->] (b) .. controls +(.5,.5) .. (c) node[midway,above] {$b_2$};
\draw[->] (c) .. controls +(-.5,-.5) .. (b) node[midway,below] {$d_2$};
\draw[->] (c) .. controls +(.5,.5) .. (d) node[midway,above] {$b_3$};
\draw[->] (d) .. controls +(-.5,-.5) .. (c) node[midway,below] {$d_3$};
\draw[->] (d) .. controls +(.5,.5) .. (e) node[midway,above] {$b_N$};
\draw[->] (e) .. controls +(-.5,-.5) .. (d) node[midway,below] {$d_N$};
\end{tikzpicture}
\caption{Birth-death process}\label{fig:bd}
\end{figure}

Birth-death processes are a subclass of Markov chains that are at the
basis, among other things, of the analysis of single-station queueing
systems (see Section~\ref{sec:single-station}). In a $(N+1)$-states
birth-death process, the transition probability (resp. rate) from
state~$i$ to~$(i+1)$ is~$b_i$, and the transition probability
(resp. rate) from state~$(i+1)$ to~$i$ is $d_i$, $i = 1, \ldots, N$
(Figure~\ref{fig:bd}). Function \verb+P = dtmcbd(b, d)+ returns a
stochastic matrix $\mathbf{P}$ for a birth-death process with birth
rates $\mathbf{b} = (b_1, \ldots, b_N)$ and death rates $\mathbf{d} =
(d_1, \ldots, d_N)$. Function \verb+Q = ctmcbd(b, d)+ does the same
for the continuous case, with the obvious difference that $\mathbf{b}$
and $\mathbf{d}$ are birth and death rates instead of probabilities.

\begin{table}[t]
\centering\begin{tabular}{lll}
\toprule
\multicolumn{2}{l}{\bf Type} & {\bf Description} \\
\cmidrule{1-2}
{\bf Continuous} & {\bf Discrete} & \\
\midrule
\verb+ctmc()+ & \verb+dtmc()+ & Stationary/Transient state occupancy probabilities \\
\verb+ctmcbd()+ & \verb+dtmcbd()+ & Birth-Death process \\
\verb+ctmcexps()+ & \verb+dtmcexps()+ & Mean Sojourn Times \\
\verb+ctmctaexps()+ & \verb+dtmctaexps()+ & Time-Averaged Mean Sojourn Tiems \\
\verb+ctmcfpt()+ & \verb+dtmcfpt()+ & First Passage Times \\
\verb+ctmcmtta()+ & \verb+dtmcmtta()+ & Mean Time to Absorption \\
\bottomrule
\end{tabular}
\caption{Functions for Markov chains analysis}\label{tab:mcfunc}
\end{table}

Table~\ref{tab:mcfunc} lists the functions that compute the
performance metrics described above for.

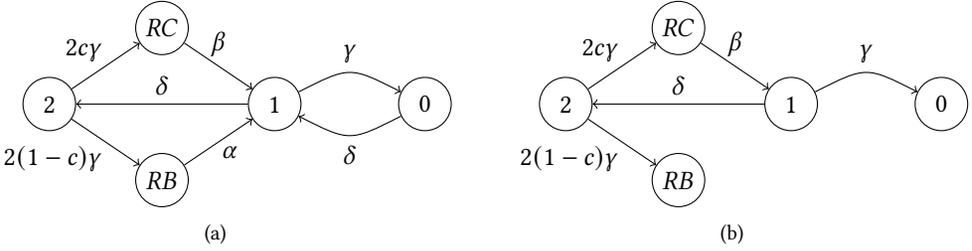
\begin{figure}[t]
  \centering
\subfigure[\label{fig:rel1}]{
\begin{tikzpicture}
\tikzstyle{every circle node} = [minimum size=7mm,draw,inner sep=0pt];
\path (0,0) node[circle] (a) {$2$}
      (3,0) node[circle] (b) {$1$}
      (5,0) node[circle] (c) {$0$}
      (1.5,1) node[circle] (rc) {$RC$}
      (1.5,-1) node[circle] (rb) {$RB$};
\draw[->] (a) -- node[above left=-2pt] {$2c\gamma$} (rc);
\draw[->] (a) -- node[below left=-2pt] {$2(1-c)\gamma$} (rb);
\draw[->] (rb) -- node[below right=-2pt] {$\alpha$} (b);
\draw[->] (rc) -- node[above] {$\beta$} (b);
\draw[->] (b) -- node[above] {$\delta$} (a);
\draw[->] (b) .. controls +(1,.5) .. (c) node[midway,above] {$\gamma$};
\draw[->] (c) .. controls +(-1,-.5) .. (b) node[midway,below] {$\delta$};
\end{tikzpicture}}\qquad%
\subfigure[\label{fig:rel2}]{
\begin{tikzpicture}
\tikzstyle{every circle node} = [minimum size=7mm,draw,inner sep=0pt];
\path (0,0) node[circle] (a) {$2$}
      (3,0) node[circle] (b) {$1$}
      (5,0) node[circle] (c) {$0$}
      (1.5,1) node[circle] (rc) {$RC$}
      (1.5,-1) node[circle] (rb) {$RB$};
\draw[->] (a) -- node[above left=-2pt] {$2c\gamma$} (rc);
\draw[->] (a) -- node[below left=-2pt] {$2(1-c)\gamma$} (rb);
\draw[->] (rc) -- node[above] {$\beta$} (b);
\draw[->] (b) -- node[above] {$\delta$} (a);
\draw[->] (b) .. controls +(1,.5) .. (c) node[midway,above] {$\gamma$};
\end{tikzpicture}}
\caption{Reliability Model for a dual-processor system (from~\cite{heiman})}\label{fig:relmodel}
\end{figure}

\subsection*{Example}
Let us consider the reliability model of a multiprocessor system shown
in Figure~\ref{fig:relmodel} and originally described
in~\cite{heiman}. The system consists of two processors, each subject
to failures with~\ac{MTTF} $1 / \gamma$. States labeled $n \in \{0, 1,
2\}$ denote that there are~$n$ working processors. If one processor
fails, it can be recovered (state \emph{RC}) with probability~$c$;
recovery takes time $1/\beta$. When the system can not be recovered, a
reboot is required (state \emph{RB}) that brings down the entire
system for time $1 / \alpha > \ 1 / \beta$. The mean time to repair a
failed processor is $1 / \delta$. The system is operational if there
is at least one working processor.

The model above can be represented as a~\ac{CTMC} with five states
$\{2, RC, RB, 1, 0\}$. The following fragment of GNU Octave code
defines the stochastic matrix $\mathbf{Q}$ of the \ac{CTMC} in
Figure~\ref{fig:rel1}, and the uses the function \verb+ctmc()+ to
compute the steady state occupancy probability vector~$\mathbf{p}$
(parameter values are taken from~\cite{heiman}):

\begin{lstlisting}
mm = 60; hh = 60*mm; dd = 24*hh; yy = 365*dd;
a = 1/(10*mm);    # 1/a = duration of reboot (10 min)
b = 1/30;         # 1/b = reconfiguration time (30 sec)
g = 1/(5000*hh);  # 1/g = processor MTTF (5000 h)
d = 1/(4*hh);     # 1/d = processor MTTR (4 h)
c = 0.9;          # recovery probability
#       2    RC      RB        1     0
Q = [ -2*g  2*c*g 2*(1-c)*g    0     0;   # 2
        0    -b       0        b     0;   # RC
        0     0      -a        a     0;   # RB
        d     0       0     -(g+d)   g;   # 1
        0     0       0        d    -d];  # 0
p = ctmc(Q);
\end{lstlisting}

\noindent that is $\mathbf{p} = (9.9839\times 10^{-1}, 2.9952\times
10^{-6}, 6.6559\times 10^{-6}, 1.5974\times 10^{-3}, 1.2779\times
10^{-6})$. From these values we can derive several availability
metrics; for example, the average time spent over one year in
states~$RC$, $RB$ and~$0$ is:

\begin{lstlisting}
p(2)*yy/mm  # minutes/year spent in RC
# => 1.5743
p(3)*yy/mm  # minutes/year spent in RB
# => 3.4984
p(5)*yy/mm  # minutes/year spent in 0
# => 0.67169
\end{lstlisting}

\noindent that is, over a year, the system is unavailable for
about~$1.57$ minutes due to reconfigurations, $3.50$ minutes due to
reboots and~$0.67$ minutes due to failure of both processors.

The~\ac{MTBF} is the average duration of continuous system
operation. We assume that the system starts in state~$2$, and we
consider the system operational also when in the reconfiguration
state. Therefore, the set of states that we consider operational is
$\{2, 1, RC\}$. If we make states~$0$ and~$\mathit{RB}$ absorbing by
removing all their outgoing transitions, the~\ac{MTBF} is the mean
time to absorption of the (modified) \ac{CTMC}:

\begin{lstlisting}
Q(3,:) = Q(5,:) = 0;      # make states {0, RB} absorbing
p0 = [1 0 0 0 0];         # initial state occupancy prob.
MTBF = ctmcmtta(Q, p0)/yy # MTBF (years)
# => 2.8376
\end{lstlisting}

\noindent that yields a~\ac{MTBF} of approximately~$2.84$ years.

\section{Single-station queueing systems} \label{sec:single-station}

A single-station queueing system, also called \emph{service center},
consists of one or more servers connected to a shared queue. An
infinite stream of requests (jobs) is generated outside the system and
put into the queue. Jobs are extracted according to some queueing
policy (e.g., First-Come-First-Served) and processed by one of the
available servers. Once service completes, a job leaves the system
permanently.

The following information is required to fully describe a
single-station queueing system: (\emph{i})~the nature of the arrival
process; (\emph{ii})~the distribution of service times;
(\emph{iii})~the number of servers; (\emph{iv})~the size of the queue;
(\emph{v})~the queueing discipline, i.e., the policy used by the
server(s) to extract requests from the queue.

Kendall's notation~\cite{Kendall:1953} can be used to specify of
queueing system. It consists of five symbols $A/S/m/K/D$, where~$A$
denotes the type of arrival process, $S$ the service time
distribution, $m \geq 1$ the number of servers, $K \geq m$ the maximum
system capacity, and~$D$ the queueing discipline.

Several types of arrival processes~$A$ and service time
distributions~$S$ have been studied in the literature, and assigned
specific symbols: $M$ (exponential distribution), $D$ (deterministic
distribution), $G$ (general distribution), $\textsc{Hyper}_k$
(hyperexponential distribution with~$k$ phases), and others.

Queueing disciplines include~\ac{FCFS}, \ac{LCFS}, \ac{SIRO},
and~\ac{PS}. In the~\ac{PS} discipline all jobs are served at the same
time (i.e., there is no queue), that is equivalent to round-robin
scheduling with infinitesimally small time slice.

A commonly used arrival process and service time distribution are the
Poisson point process and exponential distribution, respectively; both
are denoted with the letter~$M$ in Kendall's notation. Let~$A(t)$ be
the number of requests arriving at the queueing system during a time
interval of length~$t$; $A(t)$ is a Poisson point process if the
probability $\Pr\{A(t) = n\}$ that there are~$n$ arrivals is:
\begin{equation*}
\Pr\{A(t) = n\} = \frac{(\lambda t)^n}{n!} e^{-\lambda t}
\end{equation*}

\noindent where~$\lambda>0$ is the expected number of arrivals for
unit of time. It can be shown~\cite{kleinrock} that the stochastic
variable~$T$ representing the time between two successive arrivals
(interarrival time) follows an exponential distribution with mean
$1/\lambda$:
\begin{equation*}
\Pr\{T \leq t\} = 1 - e^{-\lambda t}
\end{equation*}

The inter-arrival and service time distributions of a $M/M/-$ queue
are therefore fully specified by the arrival rate~$\lambda$ of
requests and the throughput~$\mu$ of each server. $M/M/-$ systems have
the useful property that the~\ac{PMF} $\pi_k$ that there are $k \geq
0$ requests in the system\footnote{We adopt the widely used convention
  of using the same symbol~$\pi_k$ for both the state occupancy
  probability of a queueing system and the state of a Markov chain}
has a simple form allowing stationary performance measures to be
expressed easily~\cite{kleinrock,bolch}.

\begin{figure}
  \centering\includegraphics[scale=0.7]{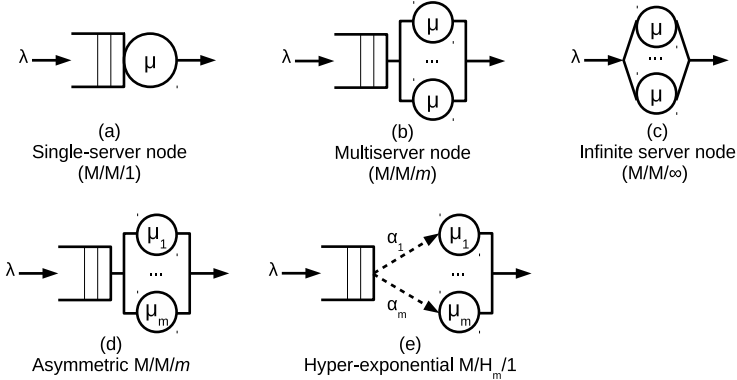}
  \caption{Graphical representation of some single-station queueing
    systems}\label{fig:single-station}
\end{figure}

Figure~\ref{fig:single-station} shows a graphical representation of
some of the single-station queueing system types supported by the
\queueing package. $M/M/m$ systems have~$m \geq 1$ identical servers,
so that up to~$m$ requests can be serviced at the same time. Once a
server becomes idle, it fetches the next request from the queue (if
any) and processes it. The system is stable, i.e., the average queue
length is finite, if $\lambda < m \mu$. Special cases of the $M/M/m$
system are the $M/M/1$ service center, where there is a single server,
and the $M/M/\infty$ center where there are infinitely many identical
servers, and therefore requests do not need to wait before receiving
service. $M/M/\infty$ stations are also called~\ac{IS} nodes or
\emph{delay centers}, since they essentially delay incoming requests
by an average duration $1 /\mu$.  \ac{IS} nodes are always stable,
irrespective of the arrival and service rates.

The $M/M/m/K$ system is a finite-capacity variants of the~$M/M/m$
queueing center. The parameter~$K \geq m$ represents the maximum
number of jobs in the system, including those being served; therefore,
there are $(K-m)$ slots in the queue. Finite-capacity centers are
always stable, since each request that tries to join a full system is
discarded.

Non-Markovian queues are used in some contexts, such as modeling of
telecommunication networks. The asymmetric~$M/M/m$ systems consists
of~$m$ exponential servers with possibly different service rates
$\boldsymbol{\mu} = (\mu_1, \ldots, \mu_m)$. At most~$m$ requests can
be served concurrently; if multiple servers are available, the next
request receives service from a randomly chosen one. This system is
stable if $\lambda < \sum_{i=1}^m \mu_i$. In the
$M/\textsc{Hyper}_m/1$ system the server has~$m$ different service
rates $\boldsymbol{\mu} = (\mu_1, \ldots, \mu_m)$ that are selected
with probabilities $\boldsymbol{\alpha} = (\alpha_1, \ldots,
\alpha_m)$, $\sum_{i=1}^m \alpha_i = 1$. Non-Markovian queueing
systems are harder to analyze; the \queueing package uses the
approximation techniques described in~\cite{kleinrock}, where both
asymmetric $M/M/m$ and $M/\textsc{Hyper}_m/1$ queues are treated as
$M/G/1$ systems.

Performance measures of single-station queueing systems include
the following quantities:

\begin{description}[style=multiline]
\item[$U$] Utilization: mean fraction of time the servers are busy. In
  general, $U \in [0, 1]$: for example, for a stable $M/M/m$ system
  the utilization is $U = \lambda / (m \mu)$. In the case of the
  $M/M/\infty$ system, $U$ is defined as the \emph{traffic intensity}
  $U = \lambda / \mu$ and can be also greater than one, since the
  system is always stable.
\item[$R$] Response time: average time spent by a request inside the
  system, i.e., the mean duration of the interval between a request
  arrival in the queue and its departure after completing service.
\item[$Q$] Mean queue length.
\item[$X$] Throughput: average number of requests that complete
  service in a unit of time. If the system is stable, then the
  throughput is equal to the arrival rate ($X = \lambda$).
\end{description}

The performance measures above can be derived from the steady-state
probability~$\pi_k$, although in most cases there are simpler
closed-form expressions that do not require the explicit computation
of~$\pi_k$. However, of particular interest is the probability~$\pi_0$
that the system is empty, and the rejection probability~$\pi_K$ for a
finite-capacity systems where at most~$K$ jobs are allowed. The
\queueing package can compute the value of~$\pi_k$ for Markovian
queues for any given~$k$.

Table~\ref{tab:queueing-systems} lists the functions provided by the
\queueing package to analyze the supported types of queueing systems.
Note that $M/M/1$ and $M/M/\infty$ systems are handled separately from
$M/M/m$ queues, since simpler formulas for the special cases $m=1$ and
$m=\infty$ are used.

\begin{table*}[t]
\centering%
\begin{tabular}{ll}
\toprule
{\bf Function} & {\bf Description}\\
\midrule
\verb+qsmm1()+   & $M/M/1$ system \\
\verb+qsmmm()+   & $M/M/m$ system with~$m$ identical servers \\
\verb+qsmminf()+ & $M/M/\infty$ system (delay center) \\
\verb+qsmm1k()+  & $M/M/1/K$ finite-capacity system \\
\verb+qsmmmk()+  & $M/M/m/K$ finite-capacity system ($K \geq m$)\\
\verb+qsammm()+  & Asymmetric $M/M/m$ \\
\verb+qsmh1()+   & $M/\textsc{Hyper}_m/1$ queue with hyper-exponential service time distribution \\
\verb+qsmg1()+   & $M/G/1$ queue with general service time distribution \\
\bottomrule
\end{tabular}
\caption{Supported single-station queueing systems}\label{tab:queueing-systems}
\end{table*}

\subsection*{Example}
Let us consider a $M/M/m$ center with arrival rate~$\lambda$ and
service rates~$\mu$. Assuming stability ($\lambda < m \mu$), the
steady state probability~$\pi_{k,M/M/m}$ that there are $k \geq 0$
requests in the system is~\cite{bolch}:
\begin{equation}
  \pi_{k, M/M/m} = \begin{cases}
    \pi_{0, M/M/m} \displaystyle\frac{(m \rho)^k}{k!} & 0 \leq k \leq m \\[4mm]
    \pi_{0, M/M/m} \displaystyle\frac{\rho^k m^m}{m!} & k > m
  \end{cases}\label{eq:pimmm}
\end{equation}

\noindent where $\rho = \lambda / (m \mu)$ is the individual server
utilization, and the steady-state probability $\pi_{0, M/M/m}$ that there are
no requests in the system is:
\begin{equation}
  \pi_{0, M/M/m} = \left[ \sum_{k=0}^{m-1} \frac{(m\rho)^k}{k!} + \frac{(m \rho)^m}{m!} \frac{1}{1 - \rho}\right]^{-1} \label{eq:pi-k-mminf}
\end{equation}

The limit of~\eqref{eq:pimmm} as~$m$ tends to infinity is the steady
state probability $\pi_{k, M/M/\infty}$ that there are~$k$ request in
a $M/M/\infty$ \acl{IS} node:
\begin{equation*}
\pi_{k, M/M/\infty} = \lim_{m \rightarrow \infty} \pi_{k, M/M/m} = \frac{1}{k!} \left( \frac{\lambda}{\mu} \right)^k e^{-\lambda / \mu}
\end{equation*}

The following fragment of GNU Octave code uses the functions
\verb+qsmmm()+ and \verb+qsmminf()+ to compute the steady state
probability that there are~$k$ requests in the system, $k=0, \ldots,
20$, for an $M/M/4$, $M/M/5$ and $M/M/\infty$ system.

\begin{lstlisting}
lambda = 4; mu = 1.2; k = 0:20;
pi_mm4   = qsmmm(lambda, mu, 4, k);
pi_mm5   = qsmmm(lambda, mu, 5, k);
pi_mminf = qsmminf(lambda, mu, k);
\end{lstlisting}

Note that \verb+qsmmm()+, \verb+qsmmm()+ and \verb+qsmminf()+ like
other functions in the \queueing package, support vector arguments.
In these cases a vector of results is returned. Also, the \queueing package
relies on Horner's rule
\begin{equation*}
\sum_{k=0}^n \displaystyle\frac{a^k}{k!} = 
  1 + a \left( 1 + \frac{a}{2} \left( 1 + \frac{a}{3} \left( \cdots \left( 1 + \frac{a}{n} \right) \cdots \right) \right) \right)
\end{equation*}
\noindent to evaluate the summation of Eq.~\eqref{eq:pi-k-mminf} more
accurately.

\begin{figure}
  \centering\includegraphics[scale=.7]{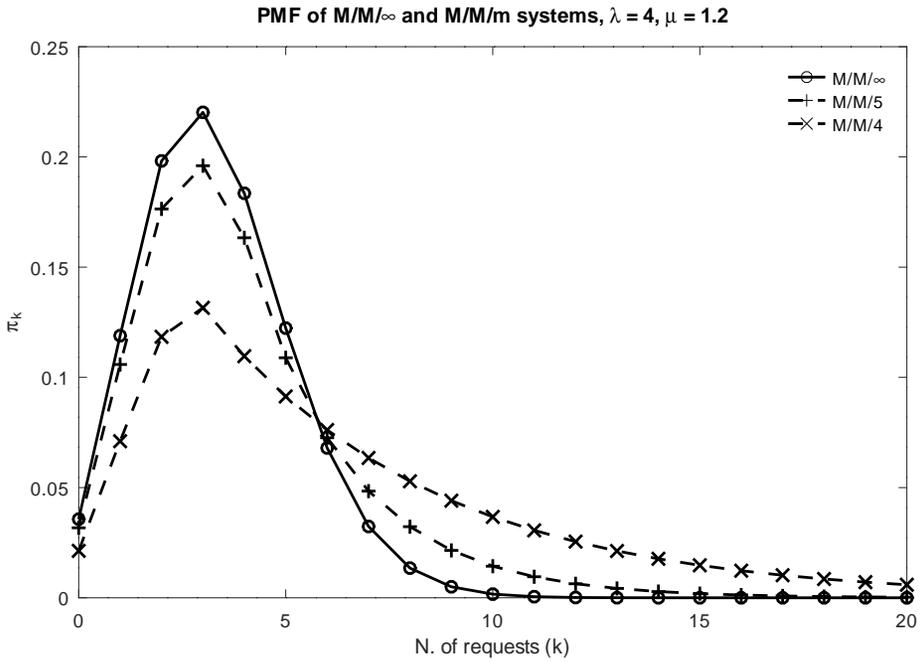}
  \caption{Probability $\pi_k$ that there are~$k$ requests in a
    $M/M/4$, $M/M/5$ and $M/M/\infty$ system with arrival rate
    $\lambda = 4\mathrm{req}/s$ and service rates $\mu = 1.2
    \mathrm{req}/s$.}\label{fig:mminf}
\end{figure}

Figure~\ref{fig:mminf} shows that the marginal probabilities $\pi_{k,
  M/M/m}$ tend indeed to $\pi_{k, M/M/\infty}$ as the number of
servers~$m$ grows.

\section{Queueing Networks}\label{sec:queueing-networks}

\begin{figure}
  \centering\includegraphics[scale=.7]{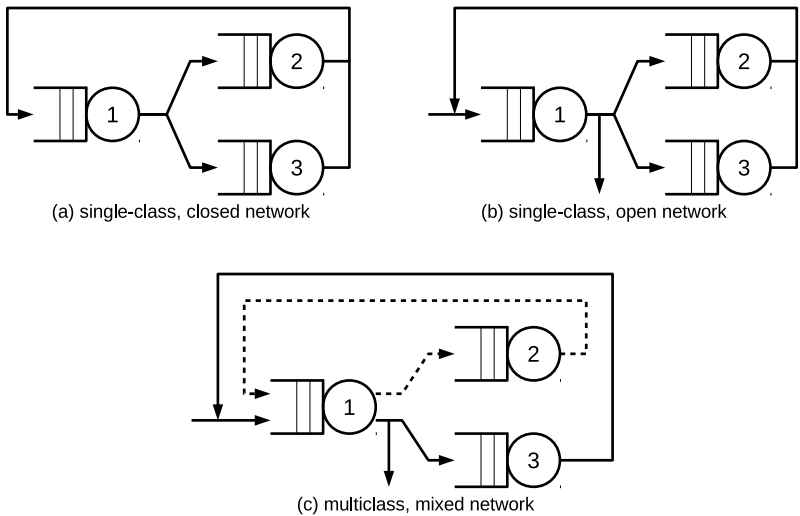}
  \caption{Three types of queueing networks: (a)~Closed network;
    (b)~Open network; (c)~Mixed network.}\label{fig:qn_types}
\end{figure}

A~\ac{QN} consists of~$K \geq 1$ service centers (nodes) and a
population of requests (jobs) that visit the servers in some
order. Several types of~\acp{QN} have been studied, depending on the
type of population of requests. In \emph{open networks} there is an
infinite stream of jobs that originate outside the system and
eventually leave the system forever (Figure~\ref{fig:qn_types}a). In
\emph{closed networks} there is a fixed population of jobs that never
leave the system (Figure~\ref{fig:qn_types}b). Requests can be all of
the same type (\emph{single-class networks}) or of multiple types
(\emph{multiclass models}). In a multiclass~\ac{QN}, different types
of requests can visit the service centers in a different order or have
different service demands (the service demand is the average time
spent by requests on a given node, see below). In mixed networks, open
and closed classes of requests can coexist
(Figure~\ref{fig:qn_types}c).

\ac{QN} analysis consists of computing steady-state performance
measures such as throughput or average queue length of the service
centers. These measures can be derived from the equilibrium state
probability~$\Pr(\mathcal{S})$ that the system is in state
$\mathcal{S}$ for each valid state, where the exact nature
of~$\mathcal{S}$ is model-dependent.

Some classes of~\acp{QN} enjoy product-form solution, meaning
that~$\Pr(\mathcal{S})$ has the relatively simple form
\begin{equation}
\Pr(\mathcal{S}) = \frac{1}{G(\mathcal{S})} d(\mathcal{S}) \prod_{i=1}^K f_i(x_i)\label{eq:product-form}
\end{equation}

\noindent where~$x_i$ is the configuration of the $i$-th service
center, $f_i$ is a function that depends on the type of service
center, $d(\mathcal{S})$ is a model-dependent function of the global
state, and~$G$ is a normalization constant. A~\ac{QN} with
product-form solution can be analyzed efficiently by considering each
node in isolation and combining the partial results.

The first class of product-form open networks was identified by
Jackson~\cite{jackson}. Later, Gordon and Newell extended product-form
solution to certain classes of closed
networks~\cite{GordonNewell}. These results were further extended by
Baskett, Chandy, Muntz and Palacios~\cite{BCMP} to include open,
closed and mixed networks with multiple customer classes (since then
known as~BCMP networks). Other types of networks have been shown to
possess product-form solution; the interested reader is referred
to~\cite{Balsamo2000} for a review.

The \queueing package supports a subset of~BCMP networks that satisfy
the following constraints:

\begin{itemize}
\item The network can consist of open or closed job classes (or both).
\item The following queueing disciplines are allowed: \ac{FCFS},
  \ac{PS}, \ac{LCFS-PR} and~\ac{IS}.
\item Service times for~\ac{FCFS} nodes are exponentially distributed
  and class-independent. For~\ac{PS}, \ac{LCFS-PR} and~\ac{IS} nodes,
  different classes of customers can have different service times.
\item The service rate of a~\ac{FCFS} node can depend on the number
  of jobs at this node (load-dependent service centers).
\item In open networks two kinds of arrival processes are allowed:
  (\emph{i})~Poisson arrival process with arrival
  rate~$\lambda$. (\emph{ii})~$C$ independent Poisson arrival streams
  where the~$C$ job sources are assigned to the~$C$ chains.
\end{itemize}

The constraints above allow a considerable simplification of the
algorithms implemented, and at the same time include the types of
networks that are most frequently used in practice.

\begin{table}[t]
\centering\small%
\begin{tabular}{llp{.65\textwidth}}
  \toprule
  \multicolumn{2}{l}{\bf Number of classes} & {\bf Description}\\
  \cmidrule{1-2} 
  {\bf Single} & {\bf Multiple} & \\
  \midrule 
\verb+qncsaba()+ & \verb+qncmaba()+ & Asymptotic Bounds for closed Networks~\cite{aba} \\
\verb+qnosaba()+ & \verb+qnomaba()+ & Asymptotic Bounds for open networks~\cite{aba} \\
\verb+qncsbsb()+ & & Balanced System Bounds for closed networks~\cite{bsb}\\
\verb+qnosbsb()+ & \verb+qncmbsb()+ & Balanced System Bounds for open networks~\cite{bsb}\\
\verb+qncsgb()+ & & Geometric Bounds~\cite{gb}\\
& \verb+qncmcb()+ & Composite Bound Method~\cite{cb}\\
\verb+qnos()+ & \verb+qnom()+ & Analysis of Jackson networks~\cite{jackson} and their multiclass extension\\
\verb+qncsconv()+ & & Convolution algorithm for closed networks with fixed-rate servers~\cite{convolution}\\
\verb+qncsconvld()+ & & Convolution algorithm for closed networks with load dependent servers~\cite{convolution}\\
\verb+qncsmva()+ & \verb+qncmmva()+ & \ac{MVA} for closed networks with fixed-rate and multiple server nodes~\cite{mva, schwetman}\\
\verb+qncsmvald()+ & & \ac{MVA} for closed networks with load dependent servers~\cite{mva}\\
\verb+qncscmva()+ & & Conditional~\ac{MVA} for closed networks with a load dependent server~\cite{cmva}\\
& \verb+qncmmvabs()+ & Bard and Schweitzer's \ac{MVA} approximation for closed networks with fixed-rate servers~\cite{bard79,scw79}\\
& \verb+qnmix()+ & \ac{MVA} for multiclass networks with both open and closed chains and fixed-rate servers~\cite{schwetman} \\
\verb+qncsmvablo()+ & & Approximate~\ac{MVA} for closed networks with blocking~\cite{mvablo}\\
\verb+qn?svisits()+ & \verb+qn?mvisits()+ & Compute visit ratios from the routing matrix (\verb+?+ can be \verb+c+ or \verb+o+)\\
\bottomrule
\end{tabular}
\caption{Main functions for~\ac{QN} analysis}\label{tab:functions}
\end{table}

Table~\ref{tab:functions} lists the main functions for~\ac{QN}
analysis provided by the \queueing package; more details are provided
in the rest of this section.

\paragraph*{Single-class models}
In single class models, service centers do not differentiate the
requests that they process. This means that, for example, the mean
time spent by a request in a given server (service time) will depend
only on the server, not on the type of request.

A single-class~\ac{QN} can be fully specified by the following
parameters:

\begin{description}[style=multiline]
\item[$K$] Number of service centers.
\item[$\lambda_i$] (Open networks only) External arrival rate to
  center~$i \in \{1, \ldots, K\}$.
\item[$N$] (Closed networks only) Total number of requests in the
  system.
\item[$Z$] (Closed networks only) Optional external delay (``think
  time'') spent by each request outside the system after each
  interaction.
\item[$S_i$] Mean service time at any server inside center~$i$ for
  each visit (not including the time spent waiting in the queue). For
  general load-dependent service centers, the service time is a vector
  where $S_i(n)$ is the service time when there are~$n$ requests in
  center~$i$
\item[$P_{i, j}$] Probability that a request completing service at
  center~$i$ is routed to center~$j$. For open networks, the
  probability that a request leaves the system after completing
  service at center~$i$ is $\left(1-\sum_{j=1}^K P_{i, j}\right)$.
\item[$V_i$] Mean number of visits to center~$i$ (also called
  \emph{visit ratio} or \emph{relative arrival rate}).
\end{description}

For open, single class networks the visit ratios~$V_i$ satisfy the
following equations:
\begin{equation}
  V_i = P_{0, i} + \sum_{j=1}^K V_j P_{j, i} \quad i=1, \ldots, K\label{eq:v-open}
\end{equation}

\noindent where $P_{0, i}$ is the probability that an external request
goes to center~$i$. If we denote with $\lambda_i$ the external arrival
rate to center~$i$, and $\lambda = \sum_i \lambda_i$ is the overall
external arrival rate, then $P_{0, i} = \lambda_i / \lambda$.

For closed networks, the visit ratios satisfy the following equation:
\begin{equation}
  \begin{cases}
    V_i = \sum_{j=1}^K V_j P_{j, i} & i=1, \ldots, K,\ i \neq r \\
    V_r = 1 & \mbox{for a selected reference station $r \in \{1, \ldots, K\}$}
  \end{cases}\label{eq:v-closed}
\end{equation}

The second condition ensures that the values~$V_i$ are uniquely
defined. A job that returns to the reference station (default $r=1$)
is assumed to have completed one interaction with the system. The
product $D_i = S_i V_i$ of the average service time per visit~$S_i$
and the mean number of visits~$V_i$ is called \emph{service demand},
and can be understood as the total service time requested by a job
during one interaction with the system. The service center with the
larger service demand is the bottleneck of the system.

Most of the algorithms in the \queueing package rely on the visit
ratios~$V_i$; if only the routing matrix~$\mathbf{P}$ is available,
functions \verb+qnosvisits()+ and \verb+qncsvisits()+ can be used to
compute the~$V_i$ using Eq.~\eqref{eq:v-open} or~\eqref{eq:v-closed},
respectively.

The following performance results for single-class models are computed:

\begin{description}[style=multiline]
\item[$U_i$] Utilization of service center~$i$;
\item[$R_i$] Response time of service center~$i$;
\item[$Q_i$] Average number of requests at center~$i$, including
  the request(s) being served;
\item[$X_i$] Throughput of service center~$i$;
\end{description}

From the values above, global performance measures can be derived:
\begin{description}[style=multiline]
\item[$X$] System throughput, $X = X_i / V_i$ for any~$i$ for which $V_i > 0$;
\item[$R$] System response time, $R = \sum_{i=1}^K R_i V_i$;
\item[$Q$] Average number of requests in the system, $Q = \sum_{i=1}^K Q_i$
\end{description}

The~\ac{MVA}~\cite{mva} and convolution~\cite{convolution} algorithms
are the most widely used techniques to compute stationary performance
measures of closed product-form networks. The convolution algorithm
computes the normalization constant~$G$ in
Eq.~\eqref{eq:product-form}; all other performance measures are
derived from~$G$. \ac{MVA} relies on the fact that, in a closed
network with~$N$ requests, the response time~$R_i(N)$ of center~$i$
can be expressed as~\cite{mva}
\begin{equation}
  R_i(N) = S_i\left(1+Q_i(N-1)\right)\label{eq:mva-recursion}
\end{equation}
\noindent where~$Q_i(N-1)$ is the mean queue length at center~$i$ if
one request is removed from the system. In the case of a single-class
network with $M/M/1$ center or $M/M/\infty$ \ac{IS} nodes only,
\ac{MVA} assumes the simple form shown in Algorithm~\ref{alg:mva}.

\begin{algorithm}[t]
  \caption{MVA algorithm without load-dependent service centers}\label{alg:mva}
  \begin{algorithmic}
    \Require $K, N, Z, S_i, V_i$, $i=1, \ldots, K$
    \Ensure $Q_i, R_i, U_i, X_i$
    \For{$i \gets 1, \ldots, K$}
    \State $Q_i \gets 0$
    \EndFor
    \For{$n \gets 1, \ldots, N$}
    \For{$i \gets 1, \ldots, K$}
    \State $R_i \gets \begin{cases} S_i & \mbox{if center $i$ is $M/M/\infty$}\\ S_i(1+Q_i) & \mbox{if center $i$ is $M/M/1$}\end{cases}$
    \EndFor
    \State $R \gets \displaystyle\sum_{i=1}^M R_i V_i$
    \State $X = \displaystyle\frac{n}{Z+R}$
    \For{$i \gets 1, \ldots, K$}
    \State $Q_i \gets X V_i R_i$
    \EndFor
    \EndFor
    \For{$i \gets 1, \ldots, K$}
    \State $X_i \gets X V_i$
    \State $U_i \gets X V_i S_i$
    \EndFor
  \end{algorithmic}
\end{algorithm}

A single-class closed networks with~$N$ requests and~$K$ service
centers of type $M/M/1$ or $M/M/\infty$ can be analyzed in
time~$O(NK)$ by either the~\ac{MVA} or the convolution
algorithms. Unfortunately, if multiple-server nodes or general
load-dependent service centers are present, both algorithms suffer
from numerical instabilities. In the case of~\ac{MVA},
Eq.~\eqref{eq:mva-recursion} is no longer sufficient to compute the
response time at center~$i$, since adding a new request may alter the
(load dependent) service time~$S_i$. It is therefore necessary to
compute the marginal probabilities~$p_i(j|n)$ that there are~$j$
requests at center~$i$, given that the total number of requests in the
system is~$n$. At the end of each iteration, the~\ac{MVA} algorithm
computes the probability $p_i(0|n)$ that center~$i$ is idle as
\begin{equation}
p_i(0|n) = 1 - \sum_{j=1}^n p_i(j|n)\label{eq:mva-error}
\end{equation}

Eq.~\eqref{eq:mva-error} is the source of numerical
errors~\cite{reiser1981}, especially if there are many requests ($n$
is large) and/or there are servers whose utilization is close
to~$1$.

So far, no numerically stable variant of the~\ac{MVA} and convolution
algorithms exist, although stable approximations have been
proposed~\cite{smva}. The \queueing package provides an implementation
of the~\ac{CMVA} algorithm~\cite{cmva}, a numerically stable variant
of~\ac{MVA}. Unfortunately, \ac{CMVA} only supports a single
load-dependent service center, and is therefore less general
than~\ac{MVA}.

\paragraph*{Multiple-class models}
The~\ac{MVA} and convolution algorithms can be extended to
handle~\acp{QN} with multiple job classes. In a multiclass~\ac{QN}
there are~$C$ customer classes; open and closed classes of requests
may be present at the same time. Since a request may change class
after service completion, the concept of \emph{chain} needs to be
introduced. Chains induce a partition of the set of classes:
class~$c_1$ and~$c_2$ belong to the same chain if a job of class~$c_1$
can eventually become a job of class~$c_2$. A chain can contain
multiple classes, but can not contain both an open and a closed class.
This prevents jobs from closed classes to enter open classes, or the
other way around

A multiclass network can be described using the same parameters as
those used for single class models, with additional subscripts
required to take classes into account:

\begin{description}[style=multiline,leftmargin=12mm]
\item[$\lambda_{c,i}$] (Open networks only) External arrival rate of
  class~$c$ requests to service center~$i$.
\item[$N_c$] (Closed networks only) Total number of class~$c$ requests
  in the system.
\item[$Z_c$] (Closed networks only) External delay (also called
  ``think time'') spent by each class~$c$ request outside the system
  after one round of interaction with the service centers is
  completes. See below.
\item[$S_{c.i}$] Mean service time of class~$c$ requests at
  center~$i$; product-form requires that service times at~\ac{FCFS}
  queues be class-independent, while service times at~\ac{IS}
  or~\ac{PS} nodes can vary on a per-class basis.
\item[$P_{r, i, s, j}$] Probability that a class~$r$ request that
  completes service at center~$i$ is routed to class~$j$ as a
  class~$s$ request.
\item[$V_{c, i}$] Mean number of visits of class~$c$ requests to
  center~$i$.
\end{description}

Similarly, performance results (utilization, response times, and so
on) are computed for each service center and class, e.g., $X_{c,i}$
denotes the throughput of class~$c$ requests at center~$i$.

\begin{algorithm}[t]
  \caption{Multiclass MVA without load-dependent service centers}\label{alg:multiclass-mva}
  \begin{algorithmic}
    \Require $K, C, N_c, Z_c, S_{c,i}, V_{c,i}$, $c=1, \ldots, C$, $i=1, \ldots, K$
    \Ensure $Q_i, R_{c,i}, X_c$
    \For{$i \gets 1, \ldots, K$}
    \State $Q_i(\mathbf{0}) \gets 0$
    \EndFor
    \For{$n \gets 1, \ldots, \sum_{c=1}^C N_c$}
    \For{each feasible population $\mathbf{n} = (n_1, \ldots, n_C)$ with~$n$ total requests}
    \For{$c \gets 1, \ldots, C$}        
    \For{$i \gets 1, \ldots, K$}
    \State $R_{c,i} \gets \begin{cases} S_{c,i} & \mbox{if center $i$ is $M/M/\infty$}\\ S_{c,i}\left(1+Q_i(\mathbf{n} - \mathbf{1}_c)\right) & \mbox{if center $i$ is $M/M/1$}\end{cases}$
    \EndFor
    \EndFor
    \For{$c \gets 1, \ldots, C$}
    \State $X_c = \displaystyle\frac{n_c}{Z_c + \sum_{i=1}^K V_{c,i} R_{c,i}}$
    \EndFor
    \For{$i \gets 1, \ldots, K$}
    \State $Q_i(\mathbf{n}) \gets \displaystyle\sum_{c=1}^C X_c V_{c,i} R_{c,i}$
    \EndFor
    \EndFor
    \EndFor
  \end{algorithmic}
\end{algorithm}

The \queueing package analyzes product-form multiclass closed networks
are using the multiclass~\ac{MVA} algorithm. Let $\mathbf{N} = (N_1,
\ldots, N_C)$ be the population vector, i.e., the vector where~$N_c$
is the number of class~$c$ requests in the system, $c=1, \ldots,
C$. Let $\mathbf{1}_c$ be the vector of length~$C$ where the $c$-th
element is one and all other elements are zero.  For closed networks
with only fixed-rate ($M/M/1$) and~\ac{IS} ($M/M/\infty$) nodes,
the~BCMP theorem~\cite{BCMP} states that the response time
$R_{c,i}(\mathbf{N})$ of class~$c$ requests at center~$i$ is:
\begin{equation}
  R_{c,i}(\mathbf{N}) = S_{c,i}\left(1+Q_i(\mathbf{N}-\mathbf{1}_c)\right)\label{eq:multiclass-mva-recursion}
\end{equation}
\noindent where $Q_i(\mathbf{N} - \mathbf{1}_c)$ is the mean queue
length at center~$i$ with one class~$c$ customer removed (if $n_c = 0$
we let $Q_i(\mathbf{N} - \mathbf{1}_c) =
0$). Eq.~\eqref{eq:multiclass-mva-recursion} is similar
to~\eqref{eq:mva-recursion}, and is the core of the
multiclass~\ac{MVA} Algorithm shown in~\ref{alg:multiclass-mva}.

Multiclass~\ac{MVA} allows all performance measures to be computed
starting from the queue lengths~$Q_i(\mathbf{0}) = 0$ of the network
with no jobs. Specifically, Algorithm~\ref{alg:multiclass-mva}
computes the mean response time $R_{c,i}$ of class~$c$ requests at
center~$i$, the mean queue length $Q_i(\mathbf{N})$ at center~$i$, and
the global throughput~$X_c$ of class~$c$ requests. The other
performance measures can be derived easily~\cite{lazowska,bolch}:
\begin{align}
  X_{c,i} &= X_c V_{c,i} && \text{class $c$ throughput at center $i$}\label{eq:multiclass-tput} \\
  U_{c,i} &= X_c S_{c,i} V_{c,i} && \text{class $c$ utilization at center $i$}\\
  Q_{c,i} &= X_c R_{c,i} && \text{mean number of class $c$ requests at center $i$}
\end{align}

The multiclass~\ac{MVA} algorithm generates all feasible populations
$\mathbf{n} = (n_1, \ldots, n_C)$; we say that $\mathbf{n}$ is
feasible with respect to the population vector $\mathbf{N} = (N_1,
\ldots, N_C)$ if $0 \leq n_c \leq N_c$ for all $c = 1, \ldots, C$. It
can be easily seen that there are $\prod_{c=1}^C (N_c+1)$ feasible
population vectors; therefore, for a closed, multiclass network
with~$K$ fixed-rate or~\ac{IS} nodes, $C$ customer classes and
population vector~$\mathbf{N}$, multiclass~\ac{MVA} requires time
$O\left(CK \prod_{c=1}^C (N_c + 1) \right)$ and space $O\left(K
\prod_{c=1}^C (N_c + 1) \right)$.

Due to its computational cost, multiclass~\ac{MVA} is appropriate for
networks with small population and limited number of classes. For
larger networks, approximations based on the~\ac{MVA} have been
proposed in the literature. The \queueing package provides an
implementation of Bard and Schweitzer's iterative approximation
scheme~\cite{bard79, scw79, lazowska} through function
\verb+qncmmvabs()+. Bard-Scweitzer approximation requires
space~$O(CK)$, that compares favorably with that of standard
multiclass~\ac{MVA}. Being an iterative scheme that stops as soon as a
convergence criterion is met, the execution time of Bard-Scweitzer
approximation depends on the network being analyzed, but is generally
much lower than multiclass~\ac{MVA} (see the example at the end of
this section). Unfortunately, the drawback is that there is no known
way to estimate the accuracy of the results provided by the
Bard-Scweitzer algorithm.

\paragraph*{Bound analysis}
In situations where accurate computation of performance measures is
impractical, bound analysis can be used to provide upper/lower limits
on the system throughput~$X$ and response time~$R$. Performance bounds
on~\acp{QN} can be computed quickly, and are useful for example in
scenarios involving on-line performance tuning of
systems~\cite{Marzolla2012,Marzolla2011,Marzolla2013}. The \queueing
package allows the computation of several classes of bounds: \ac{AB},
\ac{BSB}, \ac{CB} and~\ac{GB}.

\acl{AB}~\cite{aba} rely on the simplifying assumption that the
service demand of a request at a service center is independent from
the number of requests in the system and their exact location. Under
this assumption (that is not true in general) it is possible to bound
the system's performance by considering the extreme situations of
lowest and highest possible loads. \acp{AB} for a single-class network
with~$K$ service centers can be computed in time~$O(K)$; for
multiclass networks with~$C$ customer classes, the computational
complexity is $O(CK)$.

\acl{BSB}~\cite{bsb} provide tighter bounds that are computed by
forcing the service demands of the network under consideration to be
all the same. \ac{BSB} have the same computational complexity as
\ac{AB}, both for single and multiclass models. \acl{CB}~\cite{cb} and
\acl{GB}~\cite{gb} are yet different bounding techniques that, in many
cases, produce even better bounds with the same computational cost.

\paragraph*{Queueing networks with blocking}
The \queueing package provides limited support for analyzing closed,
single-class networks with blocking. In blocking networks, queues have
a finite capacity: a request joining a full queue will block until one
slot becomes available. Apart from very few exceptions, queueing
networks with blocking do not satisfy the conditions for product-form
solution~\cite{blocking}, and are therefore difficult to analyze.

The \verb+qncsmvablo()+ function implements the~MVABLO
algorithm~\cite{mvablo} that is based on an extension
of~\ac{MVA}. MVABLO provides approximate performance measures for
closed, single-class networks with~\ac{BAS} blocking. According to
the~\ac{BAS} discipline, a request completing service at center~$i$
that wants to move to center~$j$ blocks the source server~$i$ until
one slot is available at the destination.

Networks with a different type of blocking are handled by the
\verb+qnmarkov()+ function. This function supports single-class, open
or closed networks where all queues have (possibly different) finite
capacity. The blocking discipline is~\ac{RS-RD}: when a request
terminates service at center~$i$ and wants to move to a saturated
center~$j$, the request is put back in the queue of center~$i$ so that
it will eventually receive another round of service from~$i$.  Each
time the request completes a new round of service, it is routed to a
possibly different, randomly chosen server. In the case of open
networks, external arrivals to a saturated servers are discarded. The
\verb+qnmarkov()+ function computes performance measures by building
and analyzing the underlying Markov chain; this makes the function
unsuited for even moderate networks due to the combinatorial explosion
of the size of the Markov chain.

\subsection*{Example}

\begin{figure}[t]
  \centering\includegraphics[scale=.7]{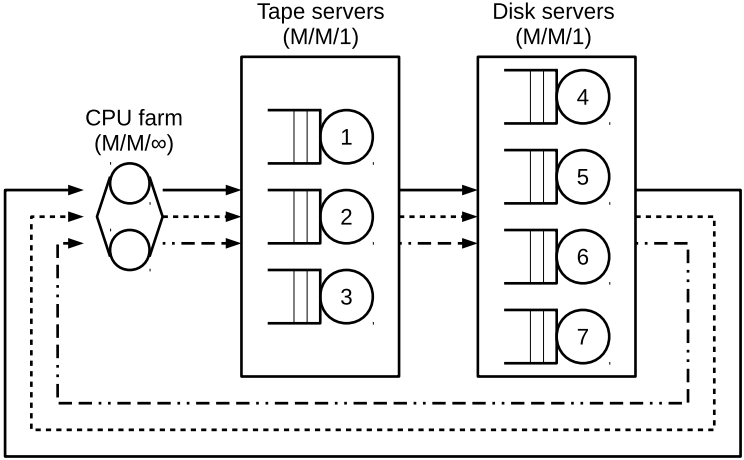}
  \caption{Multiclass closed network model of a scientific compute
    farm.}\label{fig:qn_example}
\end{figure}

We now demonstrate the use of the \queueing package for analyzing the
closed multiclass network shown in Figure~\ref{fig:qn_example}, that
represents a simple model of a scientific compute farm. The system has
three classes of jobs that process data stored on disk servers;
occasionally, data must be retrieved from tape libraries and copied to
the disk servers. Each job spends some amount of time on CPU-intensive
computations and then accesses data on external storage. Data resides
on three tape libraries (nodes 1--3). Four disk servers (nodes 4--7)
act as a cache for data copied from the tape libraries. Tape libraries
and disk servers are modeled as $M/M/1$ service centers. We assume
that the number of CPU cores is not a limiting factor, i.e., each job
starting a CPU burst always finds a CPU core available. Therefore, the
CPU farm can be represented as an~\ac{IS} node ($M/M/\infty$
center). In the model of Figure~\ref{fig:qn_example}, the~\ac{IS} node
represents the ``think time'' of jobs, a term that originated from
batch systems where~\ac{IS} nodes were the terminals where users spend
some time ``thinking'' before submitting new commands to the system.

Let~$N$ be the total number of jobs. We denote with
$\boldsymbol{\beta} = (\beta_1, \beta_2, \beta_3)$ the
\emph{population mix} of the network, where~$\beta_c$ is the fraction
of class~$c$ jobs, $0 \leq \beta_c \leq 1$ and $\beta_1 + \beta_2 +
\beta_3 = 1$. Thus, the number of class~$c$ jobs is $N_c = \beta_c N$
rounded to the nearest integer. Let~$D_{c, i}$ be the service demand
of class~$c$ requests at center~$i$ (recall that the service demand is
the product of the mean service time and the number of visits,
$D_{c,i} = S_{c,i} V_{c,i}$). Let~$Z_c$ be the average duration of a
CPU burst of a class~$c$ job. The parameter values are shown on
Table~\ref{tab:qn_example}.

\begin{table}[t]
\centering\begin{tabular}{clrrr}
\toprule
{\bf Param} & {\bf Description} & {\bf Class 1} & {\bf Class 2} & {\bf Class 3}\\
\midrule
$D_{c,1}$ & Tape Server & 100 & 180 & 280\\
$D_{c,2}$ & Tape Server & 140 & 10 & 160\\
$D_{c,3}$ & Tape Server & 200 & 70 & 150\\
$D_{c,4}$ & Disk Server & 30 & 10 & 90\\
$D_{c,5}$ & Disk Server & 50 & 90 & 20\\
$D_{c,6}$ & Disk Server & 20 & 130 & 50\\
$D_{c,7}$ & Disk Server & 10 & 30 & 18\\
$Z_c$ & Cpu farm & 2400 & 1800 & 2100\\

\bottomrule
\end{tabular}
\caption{Parameters for the model in Figure~\ref{fig:qn_example}}\label{tab:qn_example}
\end{table}

We consider $N=300$ jobs, and we want to study how different
population mixes $\boldsymbol{\beta}$ affect the system
throughput~$X$. For example, the following GNU Octave code computes
the per-class utilizations~$U_{c,i}$, response times~$R_{c,i}$, mean
queue lengths~$Q_{c,i}$ and throughput~$X_{c,i}$ when
$\boldsymbol{\beta} = (0.2, 0.3, 0.5)$:

\begin{lstlisting}
N = 300;                          # total n. of jobs
S = [100 140 200  30  50  20  10; # service demands
     180  10  70  10  90 130  30;
     280 160 150  90  20  50  18];
Z = [2400 1800 2100];             # mean duration CPU bursts
V = ones(size(S));                # n. of visits 
m = ones(1,columns(S));           # n. of servers in nodes
beta = [0.2, 0.3, 0.5];           # population mix

pop = round(N*beta); pop(3) = N - pop(1) - pop(2);
[U R Q X] = qncmmva(pop, S, V, m, Z);
X_sys = sum(X(:,1) ./ V(:,1));    # System throughput
\end{lstlisting}

Note that \verb+qncmmva()+ expects as parameters the mean service
times~$S_{c,i}$ and the mean number of visits~$V_{c,i}$. Since we know
the service demands, we let $S_{c,i} = D_{c,i}$ and set all visits to
one.

The system throughput of a multiclass network is $X_\textrm{sys} =
\sum_c X_c$, where~$X_c$ is the class~$c$ throughput. The values
of~$X_c$ can be computed from the individual servers throughput
$X_{c,i}$ that are returned by \verb+qncmmva()+, using
Eq.~\eqref{eq:multiclass-tput} with $i=1$ (actually, any valid value
for~$i$ will do). In the example above we get $X_\textrm{sys} =
0.0053793$.

Even on such a small network, \verb+qncmmva()+ requires about~$170s$
of CPU time on an Intel i7-4790 CPU at 3.60GHz running Ubuntu Linux
18.04 with GNU Octave 5.1; this makes the multiclass~\ac{MVA}
algorithm impractical for this type of study, since analyzing many
population mixes would require a prohibitive amount of time. We
therefore resort to the much faster Bard-Scweitzer approximation,
realized by function \verb+qncmmvabs()+.

\begin{figure}[t]
\centering\includegraphics[width=.7\textwidth]{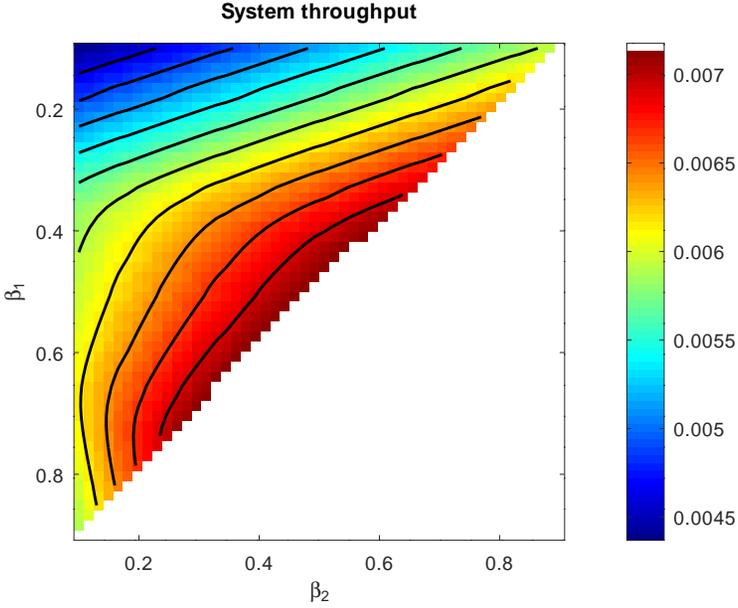}
\caption{Approximate system throughput as a function of the population
  mix $\boldsymbol{\beta} = (\beta_1, \beta_2, 1 - \beta_1 -
  \beta_2)$; the values have been computed using Bard-Schweitzer
  approximation. Contour lines show the regions of equal
  throughput. Irregularities towards the center are caused by rounding
  the population to the nearest integer. (Best viewed in
  color)}\label{fig:qn-demo}
\end{figure}

Figure~\ref{fig:qn-demo} shows the system throughput~$X$ for $(50
\times 50)/2$ different population mixes. Each square corresponds to a
combinations of $\beta_1, \beta_2$, from which $\beta_3 = N - \beta_1
- \beta_2$. Contour lines show the regions of the parameter space of
equal throughput; the population mixes that result in the optimal
throughput are those towards the center of the image.

The whole Figure~\ref{fig:qn-demo} can be computed in about~$5s$ using
\verb+qncmmvabs()+ on the same system above, i.e., orders of magnitude
faster than the time that would be required by the multiclass~\ac{MVA}
implementation from function \verb+qncmmva()+.

\section{Conclusions}\label{sec:conclusions}

In this paper we described the \queueing package, a GNU Octave package
for~\acp{QN} and Markov chains analysis. The \queueing package
includes functions for transient and stationary analysis of discrete
and continuous Markov Chains, and for stationary analysis of
single-station queueing systems and product-form~\acp{QN}. The
\queueing package can handle open, closed and mixed~\acp{QN} with one
or multiple classes of requests. Exact and approximate performance
metrics can be obtained, as well as different types of bounds.

Research on efficient solution techniques for~\ac{QN} models is still
an active topic. The \queueing package will therefore be extended to
include some of the newer algorithms. Furthermore, we plan to include
support for more types of non product-form networks.

The latest version of the \queueing package is available at
\url{https://octave.sourceforge.io/queueing/} and can be used,
modified and distributed under the terms of the GNU General Public
License (GPL) version 3 or later.

\bibliographystyle{ACM-Reference-Format}
\bibliography{octave-queueing}


\begin{thebibliography}{37}


\ifx \showCODEN    \undefined \def \showCODEN     #1{\unskip}     \fi
\ifx \showDOI      \undefined \def \showDOI       #1{#1}\fi
\ifx \showISBNx    \undefined \def \showISBNx     #1{\unskip}     \fi
\ifx \showISBNxiii \undefined \def \showISBNxiii  #1{\unskip}     \fi
\ifx \showISSN     \undefined \def \showISSN      #1{\unskip}     \fi
\ifx \showLCCN     \undefined \def \showLCCN      #1{\unskip}     \fi
\ifx \shownote     \undefined \def \shownote      #1{#1}          \fi
\ifx \showarticletitle \undefined \def \showarticletitle #1{#1}   \fi
\ifx \showURL      \undefined \def \showURL       {\relax}        \fi
\providecommand\bibfield[2]{#2}
\providecommand\bibinfo[2]{#2}
\providecommand\natexlab[1]{#1}
\providecommand\showeprint[2][]{arXiv:#2}

\bibitem[\protect\citeauthoryear{Akyildiz}{Akyildiz}{1988}]%
        {mvablo}
\bibfield{author}{\bibinfo{person}{Ian~F. Akyildiz}.}
  \bibinfo{year}{1988}\natexlab{}.
\newblock \showarticletitle{Mean Value Analysis for Blocking Queueing
  Networks}.
\newblock \bibinfo{journal}{\emph{IEEE Transactions on Software Engineering}}
  \bibinfo{volume}{1}, \bibinfo{number}{2} (\bibinfo{date}{April}
  \bibinfo{year}{1988}), \bibinfo{pages}{418--428}.
\newblock
\urldef\tempurl%
\url{https://doi.org/10.1109/32.4663}
\showDOI{\tempurl}


\bibitem[\protect\citeauthoryear{Balsamo}{Balsamo}{2000}]%
        {Balsamo2000}
\bibfield{author}{\bibinfo{person}{Simonetta Balsamo}.}
  \bibinfo{year}{2000}\natexlab{}.
\newblock \showarticletitle{Product Form Queueing Networks}.
\newblock In \bibinfo{booktitle}{\emph{Performance Evaluation: Origins and
  Directions}}, \bibfield{editor}{\bibinfo{person}{G{\"u}nter Haring},
  \bibinfo{person}{Christoph Lindemann}, {and} \bibinfo{person}{Martin Reiser}}
  (Eds.). \bibinfo{publisher}{Springer Berlin Heidelberg},
  \bibinfo{address}{Berlin, Heidelberg}, \bibinfo{pages}{377--401}.
\newblock
\showISBNx{978-3-540-46506-5}
\urldef\tempurl%
\url{https://doi.org/10.1007/3-540-46506-5_16}
\showDOI{\tempurl}


\bibitem[\protect\citeauthoryear{Balsamo, De~Nitto~Person\'e, and
  Onvural}{Balsamo et~al\mbox{.}}{2001}]%
        {blocking}
\bibfield{author}{\bibinfo{person}{S. Balsamo}, \bibinfo{person}{V.
  De~Nitto~Person\'e}, {and} \bibinfo{person}{R. Onvural}.}
  \bibinfo{year}{2001}\natexlab{}.
\newblock \bibinfo{booktitle}{\emph{Analysis of Queueing Networks with
  Blocking}}.
\newblock \bibinfo{publisher}{Kluwer Academic Publishers}.
\newblock
\showISBNx{0-7923-7996-9}


\bibitem[\protect\citeauthoryear{Bard}{Bard}{1979}]%
        {bard79}
\bibfield{author}{\bibinfo{person}{Yonathan Bard}.}
  \bibinfo{year}{1979}\natexlab{}.
\newblock \showarticletitle{Some Extensions to Multiclass Queueing Network
  Analysis}. In \bibinfo{booktitle}{\emph{Proceedings of the Third
  International Symposium on Modelling and Performance Evaluation of Computer
  Systems: Performance of Computer Systems}}. \bibinfo{publisher}{North-Holland
  Publishing Co.}, \bibinfo{address}{Amsterdam, The Netherlands, The
  Netherlands}, \bibinfo{pages}{51--62}.
\newblock
\showISBNx{0-444-85332-4}


\bibitem[\protect\citeauthoryear{Baskett, Chandy, Muntz, and Palacios}{Baskett
  et~al\mbox{.}}{1975}]%
        {BCMP}
\bibfield{author}{\bibinfo{person}{Forest Baskett}, \bibinfo{person}{K.~Mani
  Chandy}, \bibinfo{person}{Richard~R. Muntz}, {and}
  \bibinfo{person}{Fernando~G. Palacios}.} \bibinfo{year}{1975}\natexlab{}.
\newblock \showarticletitle{Open, Closed, and Mixed Networks of Queues with
  Different Classes of Customers}.
\newblock \bibinfo{journal}{\emph{J. ACM}} \bibinfo{volume}{22},
  \bibinfo{number}{2} (\bibinfo{year}{1975}), \bibinfo{pages}{248--260}.
\newblock
\showISSN{0004-5411}
\urldef\tempurl%
\url{https://doi.org/10.1145/321879.321887}
\showDOI{\tempurl}


\bibitem[\protect\citeauthoryear{Bertoli, Casale, and Serazzi}{Bertoli
  et~al\mbox{.}}{2009}]%
        {jmt}
\bibfield{author}{\bibinfo{person}{Marco Bertoli}, \bibinfo{person}{Giuliano
  Casale}, {and} \bibinfo{person}{Giuseppe Serazzi}.}
  \bibinfo{year}{2009}\natexlab{}.
\newblock \showarticletitle{{JMT}: performance engineering tools for system
  modeling}.
\newblock \bibinfo{journal}{\emph{SIGMETRICS Performance Evaluation Review}}
  \bibinfo{volume}{36}, \bibinfo{number}{4} (\bibinfo{year}{2009}),
  \bibinfo{pages}{10--15}.
\newblock
\showISSN{0163-5999}
\urldef\tempurl%
\url{https://doi.org/10.1145/1530873.1530877}
\showDOI{\tempurl}


\bibitem[\protect\citeauthoryear{Bolch, Greiner, de~Meer, and Trivedi}{Bolch
  et~al\mbox{.}}{1998}]%
        {bolch}
\bibfield{author}{\bibinfo{person}{Gunter Bolch}, \bibinfo{person}{Stefan
  Greiner}, \bibinfo{person}{Hermann de Meer}, {and} \bibinfo{person}{Kishor~S.
  Trivedi}.} \bibinfo{year}{1998}\natexlab{}.
\newblock \bibinfo{booktitle}{\emph{Queueing Networks and Markov Chains:
  Modeling and Performance Evaluation with Computer Science Applications}}.
\newblock \bibinfo{publisher}{Wiley}.
\newblock
\showISBNx{978-0-471-56525-3}


\bibitem[\protect\citeauthoryear{Buzen}{Buzen}{1973}]%
        {convolution}
\bibfield{author}{\bibinfo{person}{Jeffrey~P. Buzen}.}
  \bibinfo{year}{1973}\natexlab{}.
\newblock \showarticletitle{Computational Algorithms for Closed Queueing
  Networks with Exponential Servers}.
\newblock \bibinfo{journal}{\emph{Commun. ACM}} \bibinfo{volume}{16},
  \bibinfo{number}{9} (\bibinfo{date}{Sept.} \bibinfo{year}{1973}),
  \bibinfo{pages}{527--531}.
\newblock
\urldef\tempurl%
\url{https://doi.org/10.1145/362342.362345}
\showDOI{\tempurl}


\bibitem[\protect\citeauthoryear{Canadilla}{Canadilla}{2019}]%
        {queueing-r}
\bibfield{author}{\bibinfo{person}{Pedro Canadilla}.}
  \bibinfo{year}{2019}\natexlab{}.
\newblock \bibinfo{booktitle}{\emph{{queueing}: Analysis of Queueing Networks
  and Models}}.
\newblock
\urldef\tempurl%
\url{https://CRAN.R-project.org/package=queueing}
\showURL{%
\tempurl}
\newblock
\shownote{R package version 0.2.12.}


\bibitem[\protect\citeauthoryear{Casale}{Casale}{2008}]%
        {cmva}
\bibfield{author}{\bibinfo{person}{Giuliano Casale}.}
  \bibinfo{year}{2008}\natexlab{}.
\newblock \showarticletitle{A note on stable flow-equivalent aggregation in
  closed networks}.
\newblock \bibinfo{journal}{\emph{Queueing Syst. Theory Appl.}}
  \bibinfo{volume}{60} (\bibinfo{date}{Dec.} \bibinfo{year}{2008}),
  \bibinfo{pages}{193--202}.
\newblock
Issue 3-4.
\showISSN{0257-0130}
\urldef\tempurl%
\url{https://doi.org/10.1007/s11134-008-9093-6}
\showDOI{\tempurl}


\bibitem[\protect\citeauthoryear{Casale}{Casale}{2019}]%
        {Casale:2019}
\bibfield{author}{\bibinfo{person}{Giuliano Casale}.}
  \bibinfo{year}{2019}\natexlab{}.
\newblock \showarticletitle{Automated Multi-paradigm Analysis of Extended and
  Layered Queueing Models with LINE}. In \bibinfo{booktitle}{\emph{Companion of
  the 2019 ACM/SPEC International Conference on Performance Engineering}}
  (Mumbai, India) \emph{(\bibinfo{series}{ICPE '19})}.
  \bibinfo{publisher}{ACM}, \bibinfo{address}{New York, NY, USA},
  \bibinfo{pages}{37--38}.
\newblock
\showISBNx{978-1-4503-6286-3}
\urldef\tempurl%
\url{https://doi.org/10.1145/3302541.3311959}
\showDOI{\tempurl}


\bibitem[\protect\citeauthoryear{Casale, Gribaudo, and Serazzi}{Casale
  et~al\mbox{.}}{2011}]%
        {perfhist}
\bibfield{author}{\bibinfo{person}{Giuliano Casale}, \bibinfo{person}{Marco
  Gribaudo}, {and} \bibinfo{person}{Giuseppe Serazzi}.}
  \bibinfo{year}{2011}\natexlab{}.
\newblock \showarticletitle{Tools for Performance Evaluation of Computer
  Systems: Historical Evolution and Perspectives}.
\newblock In \bibinfo{booktitle}{\emph{Performance Evaluation of Computer and
  Communication Systems. Milestones and Future Challenges: IFIP WG 6.3/7.3
  International Workshop, PERFORM 2010, in Honor of G{\"u}nter Haring on the
  Occasion of His Emeritus Celebration, Vienna, Austria, October 14-16, 2010,
  Revised Selected Papers}}, \bibfield{editor}{\bibinfo{person}{Karin~Anna
  Hummel}, \bibinfo{person}{Helmut Hlavacs}, {and} \bibinfo{person}{Wilfried
  Gansterer}} (Eds.). \bibinfo{publisher}{Springer Berlin Heidelberg},
  \bibinfo{address}{Berlin, Heidelberg}, \bibinfo{pages}{24--37}.
\newblock
\showISBNx{978-3-642-25575-5}
\urldef\tempurl%
\url{https://doi.org/10.1007/978-3-642-25575-5_3}
\showDOI{\tempurl}


\bibitem[\protect\citeauthoryear{Casale, Muntz, and Serazzi}{Casale
  et~al\mbox{.}}{2008}]%
        {gb}
\bibfield{author}{\bibinfo{person}{Giuliano Casale}, \bibinfo{person}{R.~R.
  Muntz}, {and} \bibinfo{person}{Giuseppe Serazzi}.}
  \bibinfo{year}{2008}\natexlab{}.
\newblock \showarticletitle{Geometric Bounds: a Non-Iterative Analysis
  Technique for Closed Queueing Networks}.
\newblock \bibinfo{journal}{\emph{IEEE Trans. Comput.}} \bibinfo{volume}{57},
  \bibinfo{number}{6} (\bibinfo{date}{June} \bibinfo{year}{2008}),
  \bibinfo{pages}{780--794}.
\newblock
\urldef\tempurl%
\url{https://doi.org/10.1109/TC.2008.37}
\showDOI{\tempurl}


\bibitem[\protect\citeauthoryear{Chassell and Stallman}{Chassell and
  Stallman}{2019}]%
        {texinfo}
\bibfield{author}{\bibinfo{person}{Robert~K. Chassell} {and}
  \bibinfo{person}{Richard~M. Stallman}.} \bibinfo{year}{2019}\natexlab{}.
\newblock \bibinfo{booktitle}{\emph{Texinfo: The {GNU} Documentation Format}}.
\newblock
\showISBNx{1-882114-67-1}
\urldef\tempurl%
\url{https://www.gnu.org/software/texinfo/manual/texinfo/}
\showURL{%
\tempurl}
\newblock
\shownote{Accessed on 2013-03-13.}


\bibitem[\protect\citeauthoryear{Denning and Buzen}{Denning and Buzen}{1978}]%
        {aba}
\bibfield{author}{\bibinfo{person}{Peter~J. Denning} {and}
  \bibinfo{person}{Jeffrey~P. Buzen}.} \bibinfo{year}{1978}\natexlab{}.
\newblock \showarticletitle{The Operational Analysis of Queueing Network
  Models}.
\newblock \bibinfo{journal}{\emph{Comput. Surveys}} \bibinfo{volume}{10},
  \bibinfo{number}{3} (\bibinfo{date}{Sept.} \bibinfo{year}{1978}),
  \bibinfo{pages}{225--261}.
\newblock
\urldef\tempurl%
\url{https://doi.org/10.1145/356733.356735}
\showDOI{\tempurl}


\bibitem[\protect\citeauthoryear{Eaton, Bateman, Hauberg, and Wehbring}{Eaton
  et~al\mbox{.}}{2020}]%
        {octave}
\bibfield{author}{\bibinfo{person}{John~W. Eaton}, \bibinfo{person}{David
  Bateman}, \bibinfo{person}{S{\o}ren Hauberg}, {and} \bibinfo{person}{Rik
  Wehbring}.} \bibinfo{year}{2020}\natexlab{}.
\newblock \bibinfo{booktitle}{\emph{{GNU Octave} version 5.2.0 manual: a
  high-level interactive language for numerical computations}}.
\newblock
\urldef\tempurl%
\url{https://www.gnu.org/software/octave/doc/v5.2.0/}
\showURL{%
\tempurl}
\newblock
\shownote{Accessed on 2020-08-14.}


\bibitem[\protect\citeauthoryear{Franks, Al-Omari, Woodside, Das, and
  Derisavi}{Franks et~al\mbox{.}}{2009}]%
        {lqn}
\bibfield{author}{\bibinfo{person}{Greg Franks}, \bibinfo{person}{Tariq
  Al-Omari}, \bibinfo{person}{Murray Woodside}, \bibinfo{person}{Olivia Das},
  {and} \bibinfo{person}{Salem Derisavi}.} \bibinfo{year}{2009}\natexlab{}.
\newblock \showarticletitle{Enhanced Modeling and Solution of Layered Queueing
  Networks}.
\newblock \bibinfo{journal}{\emph{IEEE Transactions on Software Engineering}}
  \bibinfo{volume}{35}, \bibinfo{number}{2} (\bibinfo{date}{March}
  \bibinfo{year}{2009}), \bibinfo{pages}{148--161}.
\newblock
\showISSN{2326-3881}
\urldef\tempurl%
\url{https://doi.org/10.1109/TSE.2008.74}
\showDOI{\tempurl}


\bibitem[\protect\citeauthoryear{Gordon and Newell}{Gordon and Newell}{1967}]%
        {GordonNewell}
\bibfield{author}{\bibinfo{person}{William~J. Gordon} {and}
  \bibinfo{person}{Gordon~F. Newell}.} \bibinfo{year}{1967}\natexlab{}.
\newblock \showarticletitle{{Closed Queuing Systems with Exponential Servers}}.
\newblock \bibinfo{journal}{\emph{Operations Research}} \bibinfo{volume}{15},
  \bibinfo{number}{2} (\bibinfo{year}{1967}), \bibinfo{pages}{254--265}.
\newblock
\urldef\tempurl%
\url{https://doi.org/10.1287/opre.15.2.254}
\showDOI{\tempurl}


\bibitem[\protect\citeauthoryear{Gunther}{Gunther}{1997}]%
        {PDQ}
\bibfield{author}{\bibinfo{person}{Neil~J. Gunther}.}
  \bibinfo{year}{1997}\natexlab{}.
\newblock \bibinfo{booktitle}{\emph{The Practical Performance Analyst:
  Performance-by-Design Techniques for Distributed Systems}}.
\newblock \bibinfo{publisher}{McGraw-Hill, Inc.}, \bibinfo{address}{New York,
  NY, USA}.
\newblock
\showISBNx{0079129463}


\bibitem[\protect\citeauthoryear{Heiman, Mittal, and Trivedi}{Heiman
  et~al\mbox{.}}{1991}]%
        {heiman}
\bibfield{author}{\bibinfo{person}{David~I. Heiman}, \bibinfo{person}{Nitin
  Mittal}, {and} \bibinfo{person}{Kishor~S. Trivedi}.}
  \bibinfo{year}{1991}\natexlab{}.
\newblock \showarticletitle{Dependability modeling for computer systems}. In
  \bibinfo{booktitle}{\emph{Proc. Ann. Reliability and Maintainability
  Symposium}}. \bibinfo{pages}{120--128}.
\newblock
\urldef\tempurl%
\url{https://doi.org/10.1109/ARMS.1991.154425}
\showDOI{\tempurl}


\bibitem[\protect\citeauthoryear{Jackson}{Jackson}{2004}]%
        {jackson}
\bibfield{author}{\bibinfo{person}{James~R. Jackson}.}
  \bibinfo{year}{2004}\natexlab{}.
\newblock \showarticletitle{Jobshop-Like Queueing Systems}.
\newblock \bibinfo{journal}{\emph{Manage. Sci.}} \bibinfo{volume}{50},
  \bibinfo{number}{12 Supplement} (\bibinfo{year}{2004}),
  \bibinfo{pages}{1796--1802}.
\newblock
\showISSN{0025-1909}
\urldef\tempurl%
\url{https://doi.org/10.1287/mnsc.1040.0268}
\showDOI{\tempurl}


\bibitem[\protect\citeauthoryear{Kendall}{Kendall}{1953}]%
        {Kendall:1953}
\bibfield{author}{\bibinfo{person}{David~G. Kendall}.}
  \bibinfo{year}{1953}\natexlab{}.
\newblock \showarticletitle{Stochastic Processes Occurring in the Theory of
  Queues and their Analysis by the Method of the Imbedded Markov Chain}.
\newblock \bibinfo{journal}{\emph{Ann. Math. Statist.}} \bibinfo{volume}{24},
  \bibinfo{number}{3} (\bibinfo{date}{Sept.} \bibinfo{year}{1953}),
  \bibinfo{pages}{338--354}.
\newblock
\urldef\tempurl%
\url{https://doi.org/10.1214/aoms/1177728975}
\showDOI{\tempurl}


\bibitem[\protect\citeauthoryear{Kerola}{Kerola}{1986}]%
        {cb}
\bibfield{author}{\bibinfo{person}{Teemu Kerola}.}
  \bibinfo{year}{1986}\natexlab{}.
\newblock \showarticletitle{The composite bound method for computing throughput
  bounds in multiple class environments}.
\newblock \bibinfo{journal}{\emph{Performance Evaluation}} \bibinfo{volume}{6},
  \bibinfo{number}{1} (\bibinfo{date}{March} \bibinfo{year}{1986}),
  \bibinfo{pages}{1--9}.
\newblock
\showISSN{0166-5316}
\urldef\tempurl%
\url{https://doi.org/10.1016/0166-5316(86)90002-7}
\showDOI{\tempurl}


\bibitem[\protect\citeauthoryear{Kleinrock}{Kleinrock}{1975}]%
        {kleinrock}
\bibfield{author}{\bibinfo{person}{Leonard Kleinrock}.}
  \bibinfo{year}{1975}\natexlab{}.
\newblock \bibinfo{booktitle}{\emph{{Queueing Systems: Volume I--Theory}}}.
\newblock \bibinfo{publisher}{Wiley Interscience}, \bibinfo{address}{New York}.
\newblock
\showISBNx{978-0471491101}


\bibitem[\protect\citeauthoryear{Lazowska, Zahorjan, Graham, and
  Sevcik}{Lazowska et~al\mbox{.}}{1984}]%
        {lazowska}
\bibfield{author}{\bibinfo{person}{Edward~D. Lazowska}, \bibinfo{person}{John
  Zahorjan}, \bibinfo{person}{G.~Scott Graham}, {and}
  \bibinfo{person}{Kenneth~C. Sevcik}.} \bibinfo{year}{1984}\natexlab{}.
\newblock \bibinfo{booktitle}{\emph{Quantitative System Performance: Computer
  System Analysis Using Queueing Network Models}}.
\newblock \bibinfo{publisher}{Prentice Hall}.
\newblock


\bibitem[\protect\citeauthoryear{Marzolla, Ferretti, and D'Angelo}{Marzolla
  et~al\mbox{.}}{2012}]%
        {Marzolla2012}
\bibfield{author}{\bibinfo{person}{Moreno Marzolla}, \bibinfo{person}{Stefano
  Ferretti}, {and} \bibinfo{person}{Gabriele D'Angelo}.}
  \bibinfo{year}{2012}\natexlab{}.
\newblock \showarticletitle{Dynamic Resource Provisioning for Cloud-Based
  Gaming Infrastructures}.
\newblock \bibinfo{journal}{\emph{Comput. Entertain.}} \bibinfo{volume}{10},
  \bibinfo{number}{1} (\bibinfo{date}{Dec.} \bibinfo{year}{2012}).
\newblock
\urldef\tempurl%
\url{https://doi.org/10.1145/2381876.2381880}
\showDOI{\tempurl}


\bibitem[\protect\citeauthoryear{Marzolla and Mirandola}{Marzolla and
  Mirandola}{2011}]%
        {Marzolla2011}
\bibfield{author}{\bibinfo{person}{Moreno Marzolla} {and}
  \bibinfo{person}{Raffaela Mirandola}.} \bibinfo{year}{2011}\natexlab{}.
\newblock \showarticletitle{{PARSY}: Performance Aware Reconfiguration of
  Software Systems}.
\newblock \bibinfo{journal}{\emph{International Journal of Performability
  Engineering}} \bibinfo{volume}{7}, \bibinfo{number}{5}, Article
  \bibinfo{articleno}{479} (\bibinfo{year}{2011}),
  \bibinfo{numpages}{13}~pages.
\newblock
\urldef\tempurl%
\url{https://doi.org/10.23940/ijpe.11.5.p479.mag}
\showDOI{\tempurl}


\bibitem[\protect\citeauthoryear{Marzolla and Mirandola}{Marzolla and
  Mirandola}{2013}]%
        {Marzolla2013}
\bibfield{author}{\bibinfo{person}{Moreno Marzolla} {and}
  \bibinfo{person}{Raffaela Mirandola}.} \bibinfo{year}{2013}\natexlab{}.
\newblock \showarticletitle{Dynamic power management for QoS-aware
  applications}.
\newblock \bibinfo{journal}{\emph{Sustainable Computing: Informatics and
  Systems}} \bibinfo{volume}{3}, \bibinfo{number}{4} (\bibinfo{year}{2013}),
  \bibinfo{pages}{231--248}.
\newblock
\showISSN{2210-5379}
\urldef\tempurl%
\url{https://doi.org/10.1016/j.suscom.2013.02.001}
\showDOI{\tempurl}


\bibitem[\protect\citeauthoryear{{R Core Team}}{{R Core Team}}{2018}]%
        {r-lang}
\bibfield{author}{\bibinfo{person}{{R Core Team}}.}
  \bibinfo{year}{2018}\natexlab{}.
\newblock \bibinfo{booktitle}{\emph{R: A Language and Environment for
  Statistical Computing}}.
\newblock R Foundation for Statistical Computing, Vienna, Austria.
\newblock
\urldef\tempurl%
\url{https://www.R-project.org/}
\showURL{%
\tempurl}
\newblock
\shownote{Accessed on 2020-03-20.}


\bibitem[\protect\citeauthoryear{Reiser}{Reiser}{1981}]%
        {reiser1981}
\bibfield{author}{\bibinfo{person}{Martin Reiser}.}
  \bibinfo{year}{1981}\natexlab{}.
\newblock \showarticletitle{Mean-value analysis and convolution method for
  queue-dependent servers in closed queueing networks}.
\newblock \bibinfo{journal}{\emph{Performance Evaluation}} \bibinfo{volume}{1},
  \bibinfo{number}{1} (\bibinfo{year}{1981}), \bibinfo{pages}{7--18}.
\newblock
\showISSN{0166-5316}
\urldef\tempurl%
\url{https://doi.org/10.1016/0166-5316(81)90040-7}
\showDOI{\tempurl}


\bibitem[\protect\citeauthoryear{Reiser and Lavenberg}{Reiser and
  Lavenberg}{1980}]%
        {mva}
\bibfield{author}{\bibinfo{person}{Martin Reiser} {and}
  \bibinfo{person}{Stephen~S. Lavenberg}.} \bibinfo{year}{1980}\natexlab{}.
\newblock \showarticletitle{Mean-Value Analysis of Closed Multichain Queuing
  Networks}.
\newblock \bibinfo{journal}{\emph{J. ACM}} \bibinfo{volume}{27},
  \bibinfo{number}{2} (\bibinfo{date}{April} \bibinfo{year}{1980}),
  \bibinfo{pages}{313–--322}.
\newblock
\showISSN{0004-5411}
\urldef\tempurl%
\url{https://doi.org/10.1145/322186.322195}
\showDOI{\tempurl}


\bibitem[\protect\citeauthoryear{Schweitzer}{Schweitzer}{1979}]%
        {scw79}
\bibfield{author}{\bibinfo{person}{Paul~J. Schweitzer}.}
  \bibinfo{year}{1979}\natexlab{}.
\newblock \showarticletitle{Approximate Analysis of Multiclass Closed Networks
  of Queues}. In \bibinfo{booktitle}{\emph{Proc. Int. Conf. on Stochastic
  Control and Optimization}}.
\newblock


\bibitem[\protect\citeauthoryear{Schwetman}{Schwetman}{1980}]%
        {schwetman-testing}
\bibfield{author}{\bibinfo{person}{Herb Schwetman}.}
  \bibinfo{year}{1980}\natexlab{}.
\newblock \bibinfo{booktitle}{\emph{Testing Network-Of-Queues Software}}.
\newblock \bibinfo{type}{Technical Report} CSD-TR-330.
  \bibinfo{institution}{Purdue University}.
\newblock
\urldef\tempurl%
\url{https://docs.lib.purdue.edu/cstech/259/}
\showURL{%
\tempurl}
\newblock
\shownote{Accessed on 2020-03-13.}


\bibitem[\protect\citeauthoryear{Schwetman}{Schwetman}{1982}]%
        {schwetman}
\bibfield{author}{\bibinfo{person}{Herb Schwetman}.}
  \bibinfo{year}{1982}\natexlab{}.
\newblock \bibinfo{booktitle}{\emph{Implementing the Mean Value Algorithm for
  the Solution of Queueing Network Models}}.
\newblock \bibinfo{type}{Technical Report} CSD-TR-355.
  \bibinfo{institution}{Purdue University}.
\newblock
\urldef\tempurl%
\url{https://docs.lib.purdue.edu/cstech/286/}
\showURL{%
\tempurl}
\newblock
\shownote{Accessed on 2020-03-13.}


\bibitem[\protect\citeauthoryear{Trivedi and Sahner}{Trivedi and
  Sahner}{2009}]%
        {sharpe}
\bibfield{author}{\bibinfo{person}{Kishor~S. Trivedi} {and}
  \bibinfo{person}{Robin Sahner}.} \bibinfo{year}{2009}\natexlab{}.
\newblock \showarticletitle{SHARPE at the Age of Twenty Two}.
\newblock \bibinfo{journal}{\emph{SIGMETRICS Perform. Eval. Rev.}}
  \bibinfo{volume}{36}, \bibinfo{number}{4} (\bibinfo{date}{March}
  \bibinfo{year}{2009}), \bibinfo{pages}{52–57}.
\newblock
\showISSN{0163-5999}
\urldef\tempurl%
\url{https://doi.org/10.1145/1530873.1530884}
\showDOI{\tempurl}


\bibitem[\protect\citeauthoryear{Zahorjan, Sevcick, Eager, and Galler}{Zahorjan
  et~al\mbox{.}}{1982}]%
        {bsb}
\bibfield{author}{\bibinfo{person}{John Zahorjan}, \bibinfo{person}{Kenneth~C.
  Sevcick}, \bibinfo{person}{Derek~L. Eager}, {and} \bibinfo{person}{Bruce~I.
  Galler}.} \bibinfo{year}{1982}\natexlab{}.
\newblock \showarticletitle{Balanced Job Bound Analysis of Queueing Networks}.
\newblock \bibinfo{journal}{\emph{Commun. ACM}} \bibinfo{volume}{25},
  \bibinfo{number}{2} (\bibinfo{date}{Feb.} \bibinfo{year}{1982}),
  \bibinfo{pages}{134--141}.
\newblock
\urldef\tempurl%
\url{https://doi.org/10.1145/358396.358447}
\showDOI{\tempurl}


\bibitem[\protect\citeauthoryear{Zhang and Down}{Zhang and Down}{2019}]%
        {smva}
\bibfield{author}{\bibinfo{person}{Lei Zhang} {and} \bibinfo{person}{Douglas~G.
  Down}.} \bibinfo{year}{2019}\natexlab{}.
\newblock \bibinfo{booktitle}{\emph{SMVA: A Stable Mean Value Analysis
  Algorithm for Closed Systems with Load-Dependent Queues}}.
\newblock \bibinfo{publisher}{Springer International Publishing},
  \bibinfo{address}{Cham}, \bibinfo{pages}{11--28}.
\newblock
\showISBNx{978-3-319-92378-9}
\urldef\tempurl%
\url{https://doi.org/10.1007/978-3-319-92378-9_2}
\showDOI{\tempurl}


\end{thebibliography}

\end{document}